\documentclass[aps,pra,reprint,superscriptaddress,amsmath,amssymb]{revtex4-2}

\usepackage{graphicx}
\usepackage{booktabs}
\usepackage{bm}
\usepackage[colorlinks=true,linkcolor=blue,citecolor=blue,urlcolor=blue]{hyperref}

\newcommand{\neff}{n_{\mathrm{eff}}}
\newcommand{\LB}{L_{B}}
\newcommand{\Fp}{F_{+}^{(c)}}
\newcommand{\um}{\,$\mu$m}

\begin{document}

\title{On-chip topological orbital angular momentum modes at a multimode interface}

\author{Xuyang Li}
\email{xuyangli@engineering.upenn.edu}
\affiliation{Department of Materials Science and Engineering, University of Pennsylvania, Philadelphia, Pennsylvania 19104, USA}

\date{\today}

\begin{abstract}
Orbital angular momentum (OAM) offers a modal degree of freedom for integrated photonics, but an OAM waveguide coupled to its neighbors tends to lose both its confinement and its charge. We propose a multimode interface between two Su--Schrieffer--Heeger (SSH) waveguide arrays, in which a hollow waveguide supporting a pair of opposite-charge OAM modes replaces the central single-mode waveguide. When the two OAM modes couple identically to the array, without direct mixing, and are matched to it in propagation constant, the interface supports two degenerate localized states, each with a definite OAM charge in the hollow waveguide. Each state is a superposition of the topological interface mode and a dark mode of the hollow waveguide, held degenerate by chiral symmetry. Bond disorder leaves this degeneracy exact; in a disordered coupled-mode ensemble, on average 98\% of the power in the hollow waveguide remains in the launched charge after long propagation. Comparison with uniform, strongly coupled, and detuned arrays shows that, at the present interface, localization alone does not preserve the charge: degeneracy with the dark mode is also required. Mode and propagation calculations for a physical array show confined $l=\pm1$ vortex fields that keep their phase winding over 15~mm, where the launched charge still carries 86\% of the power in the hollow-waveguide region. These results provide a design route for integrating OAM-supporting waveguides into coupled photonic structures.
\end{abstract}

\maketitle

\section{Introduction}
\label{sec:intro}

Optical orbital angular momentum (OAM) adds an angular degree of freedom to amplitude, phase, polarization, and wavelength for encoding and manipulating photonic information. Advances in vortex generation~\cite{Huang2020,Miao2016,Wang2020BIC,Zhang2020,Chang2024,Hwang2024}, transmission~\cite{WangHY2023,WangJ2023}, and information processing~\cite{Devlin2017,Wu2024}, together with on-chip demonstrations using hollow waveguides~\cite{Chen2018,Chen2020} and OAM charge conversion in coupled waveguides~\cite{Qi2024}, motivate integrated structures that guide OAM fields while retaining their vortex character. A waveguide that supports OAM modes in isolation, however, need not keep them once it is coupled to other waveguides: coupling can spread the field into the neighbors and convert one OAM charge into the other. Detuning the waveguide from its neighbors suppresses both effects, but only by decoupling it from the array, which gives up the phase-matched coupling that couplers and mode converters rely on. Integrating OAM waveguides into coupled arrays therefore requires an environment that preserves both the confinement and the modal composition of the OAM field while the waveguide remains phase-matched to its neighbors.

Photonic topology provides a framework for confining light at boundaries and interfaces~\cite{Lustig2019,YangB2018,YangYH2019,Kruk2019,WangZY2023,Arkhipova2023}. In the Su--Schrieffer--Heeger (SSH) model, alternating couplings open a band gap, and an interface between chains with different winding numbers supports a localized mode, as observed in photonic waveguide, plasmonic, and nanocavity arrays~\cite{Weimann2017,Cheng2015,Han2019} and with cold atoms~\cite{Meier2016}. SSH-type models have also been realized in electronic, circuit, acoustic, and synthetic-dimension systems~\cite{Geng2022,Yue2022,DiVentra2022,Hu2023,Kanungo2022,LiGZ2023}, and related boundary and interface states have been studied in nonlinear lattices, in their dynamics, and in flat-band photonic crystals~\cite{Hadad2016,He2016,WangD2026}. In a conventional single-orbital SSH junction, one localized orbital is retained per waveguide and the domain wall supports a single topological interface state. A central OAM doublet introduces two local orbitals at the same interface site. Moreover, the existence of a localized mode alone does not ensure that an OAM superposition retains its relative phase during propagation.

One route to multiple boundary modes is to increase the bulk winding number through longer-range couplings~\cite{LiLH2015,LiuH2023}, but the required coupling hierarchy is difficult to achieve in waveguide arrays dominated by nearest-neighbor overlap. Here we instead enlarge the local mode space at the interface while retaining the surrounding SSH chains: a hollow waveguide that supports multiple local orbitals replaces the central single-mode waveguide [Figs.~\ref{fig:1}(a) and \ref{fig:1}(b)]. We refer to the single-mode waveguides as Gaussian waveguides, after the profile of their fundamental mode.

Recent studies have examined the relation between vortex localization and OAM retention~\cite{HuZ2025} and on-chip OAM protection using next-nearest-neighbor coupling~\cite{ChenC2025}. Here the surrounding array has only nearest-neighbor coupling, and the OAM degrees of freedom enter only through the local mode space of the hollow waveguide. The SSH winding number characterizes the bulk coupling structure, whereas the OAM charge describes the azimuthal phase winding of the optical field. Interface localization and retention of a definite central OAM charge are therefore separate requirements: the latter also depends on the modal propagation constants and coupling symmetry. We show that the charge is retained when the localized interface mode is degenerate with a dark mode of the hollow waveguide, and that localization without this degeneracy is not sufficient.

We describe the structure in a basis of local Gaussian and OAM modes and derive the conditions under which degenerate interface eigenstates carry a definite OAM charge in the hollow waveguide. A bright--dark decomposition shows that each state combines a topological bright interface mode with a dark mode of the hollow waveguide, held at the same propagation constant by chiral symmetry. Comparisons with isolated, uniformly coupled, strongly coupled and detuned references identify what the SSH environment contributes, and waveguide-mode and propagation simulations examine the physical array. Table~\ref{tab:platforms} places the construction alongside selected one-dimensional topological arrays in waveguide, plasmonic, and photonic-crystal platforms~\cite{Weimann2017,Cheng2015,Han2019,Song2020,WangYH2020,Pocock2018,Yan2021,Yuan2022} (Supplemental Material~\cite{SM}, Sec.~S1).

\begin{table*}[t]
\caption{Selected one-dimensional topological arrays with localized boundary states and the multimode interface considered here. In each cited system, the boundary site carries the same local mode as the bulk. For the ideal model with matched onsite propagation constants and identical coupling profiles, the interface supports one bright SSH zero mode and $N-1$ central dark modes.}
\label{tab:platforms}
\begin{ruledtabular}
\begin{tabular}{p{0.11\textwidth}p{0.22\textwidth}p{0.27\textwidth}p{0.32\textwidth}}
Refs. & Platform & Localized states studied & Local mode basis at boundary \\
\colrule
\cite{Song2020,WangYH2020} & Waveguide array & Coupled edge or interface states & Guided mode, as in bulk \\
\cite{Weimann2017} & $\mathcal{PT}$-symmetric waveguide lattice & Interface state & Guided mode, as in bulk \\
\cite{Cheng2015} & Plasmonic waveguide array & Kink and antikink interface states & SPP mode, as in bulk \\
\cite{Pocock2018,Yan2021} & Plasmonic nanoparticle chain & Edge states & Dipolar resonance, as in bulk \\
\cite{Han2019,Yuan2022} & Photonic-crystal nanocavity array & Edge states; kink interface state~\cite{Han2019} & Cavity mode, as in bulk \\
\textbf{This work} & \textbf{Waveguide array} & \textbf{Degenerate interface subspace ($\bm{N}$ zero modes, ideal model)} & \textbf{Gaussian mode in bulk; multiple central orbitals, including an OAM doublet} \\
\end{tabular}
\end{ruledtabular}
\end{table*}

\section{Theory}
\label{sec:theory}

\subsection{Coupled-mode description of the SSH chain}
\label{sec:cmt}

We describe propagation in the waveguide array using coupled-mode theory. For a fixed polarization, the optical field is expanded in localized transverse modes, and their amplitudes evolve along the propagation coordinate $z$. In the paraxial, weak-coupling approximation, the amplitudes satisfy
\begin{equation}
-i\frac{\partial a_{m,i}}{\partial z} = \beta_{m,i}\,a_{m,i} + \sum_{(m',i')\neq(m,i)} g_{(m,i),(m',i')}\,a_{m',i'}.
\label{eq:cmt}
\end{equation}
Here $a_{m,i}$ is the amplitude of mode $i$ in waveguide $m$, $\beta_{m,i}$ is its propagation constant, and $g_{(m,i),(m',i')}$ is the coupling from mode $(m',i')$ to mode $(m,i)$. The sign convention corresponds to propagation phases $\exp(i\beta z)$. This equation has the structure of a Schr\"odinger evolution equation, with propagation distance playing the role of time and propagation constants playing the role of energies. Waveguide dimensions and separations control the onsite propagation constants and coupling coefficients, respectively, through the transverse mode profiles and their overlap.

In the bulk, we retain one Gaussian mode per waveguide for each polarization and group neighboring waveguides into two-site unit cells. After subtracting a common propagation constant $\beta_0$ and retaining nearest-neighbor couplings, the Bloch Hamiltonian is
\begin{equation*}
H(k) = \begin{pmatrix} 0 & \eta^{*}(k) \\ \eta(k) & 0 \end{pmatrix}, \qquad \eta(k) = a + b\,e^{ik}.
\end{equation*}
The real positive coefficients $a$ and $b$ denote intracell and intercell couplings, respectively, and $k$ is the dimensionless Bloch wave number. Unequal couplings open a gap in the propagation-constant spectrum. The absence of onsite detuning and same-sublattice coupling gives the chiral symmetry $\sigma_z H(k)\sigma_z = -H(k)$. Within this symmetry class, the winding number
\begin{equation*}
w = \frac{1}{2\pi}\oint d\arg\eta(k)
\end{equation*}
equals one for $b>a$ and zero for $a>b$. A transition between these phases requires closing the gap as long as chiral symmetry is preserved.

We join two SSH chains with $w_{\mathrm{L}}=0$ and $w_{\mathrm{R}}=1$, as illustrated in Fig.~\ref{fig:1}(c). The left and right bulk couplings are denoted by $(a,b)$ and $(a',b')$, while $\kappa_1$ and $\kappa_2$ couple the central site to its left and right neighbors. The unit cells on both sides are chosen with the same sublattice ordering [dashed boxes in Fig.~\ref{fig:1}(c)], so that the coupling $a$ next to the interface is intracell on the left, whereas $a'$ is intercell on the right. For $a>b$ and $a'>b'$, the two chains therefore have winding numbers 0 and 1. The winding-number difference $\Delta w = w_{\mathrm{R}} - w_{\mathrm{L}} = 1$ gives the single interface mode of the conventional junction, which serves as the reference for the multimode interface below. The finite-chain Hamiltonian is given in Supplemental Material Sec.~S2.2~\cite{SM}.

\subsection{Coupling between Gaussian and OAM modes}
\label{sec:coupling}

Consider a Gaussian mode in a single-mode waveguide, WG-a, coupled to two opposite-charge OAM modes in a hollow waveguide, WG-b, as shown in inset (ii) of Fig.~\ref{fig:1}(d). Projecting the Helmholtz equation onto these three modes gives the coupled-mode Hamiltonian~\cite{Chen2020}
\begin{equation}
H = \begin{pmatrix}
\beta^{a}_{\mathrm{G}} & \kappa_{l,\mathrm{G}} & \kappa_{-l,\mathrm{G}} \\
\kappa_{\mathrm{G},l} & \beta^{b}_{l} & h_{-l,l} \\
\kappa_{\mathrm{G},-l} & h_{l,-l} & \beta^{b}_{-l}
\end{pmatrix}.
\label{eq:three}
\end{equation}
The basis consists of the Gaussian mode followed by the OAM modes with charges $+l$ and $-l$. The diagonal entries are their propagation constants, the coefficients $\kappa$ describe Gaussian--OAM coupling, and the coefficients $h$ describe mixing within the OAM pair. The derivation and normalization conventions are given in Supplemental Material Sec.~S2.1~\cite{SM}.

Hermiticity requires reverse couplings to be complex conjugates. Equal coupling to opposite OAM charges is an additional condition set by the geometry and modal symmetry: in a collinear Gaussian--annular arrangement with a reflection-symmetric scalar coupling operator, consistent choices of modal phase make the two coupling amplitudes equal. OAM-dependent coupling phases in waveguide systems have also been studied in Ref.~\cite{Turpin2017}. Equal coupling to the two charges, matched onsite propagation constants, and no direct OAM mixing define the ideal model used below; Sec.~\ref{sec:confinement} examines which departures from this model the OAM interface states tolerate.

\subsection{OAM interface eigenstates}
\label{sec:eigenstates}

\begin{figure*}[t]
\includegraphics[width=\textwidth]{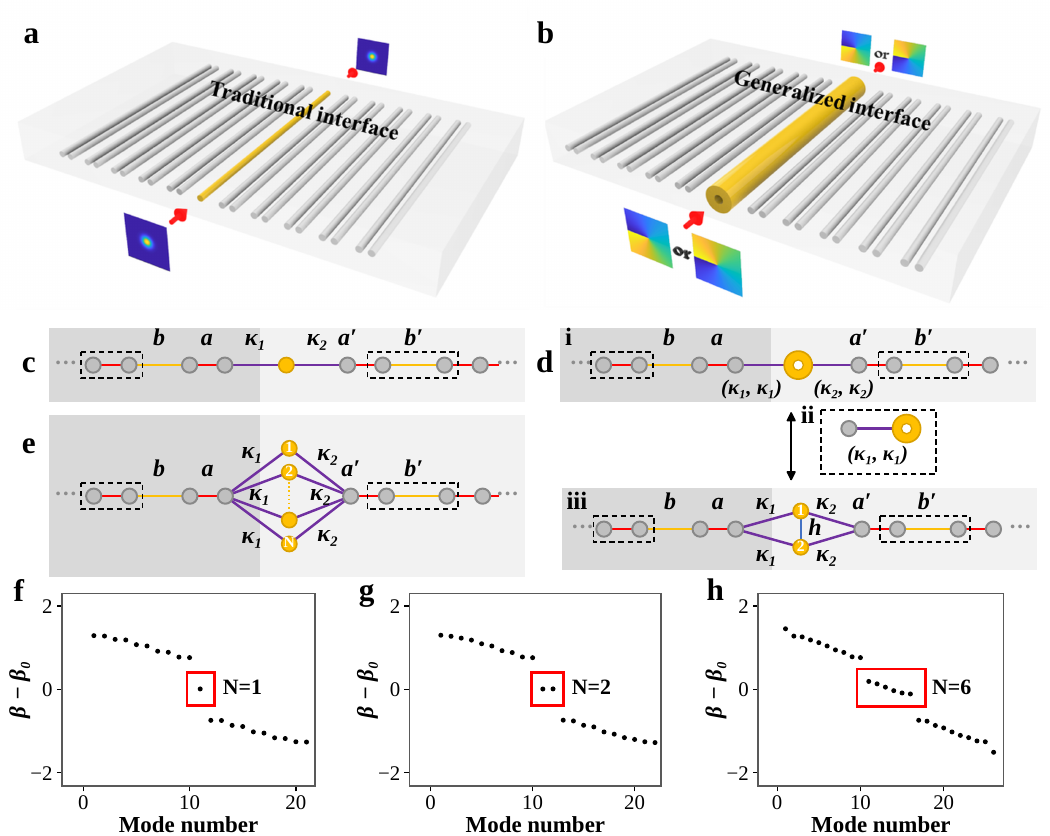}
\caption{Single-orbital and multimode interfaces in an SSH waveguide array. (a), (b) Conventional and hollow-waveguide interfaces between two SSH arrays. (c)--(e) Effective coupling models with one central orbital, two central OAM orbitals, and $N$ central orbitals, respectively. The insets in (d) illustrate the local modes and their coupling. (f)--(h) Calculated waveguide-mode spectra for the conventional interface, the two-orbital interface, and the $N=6$ interface; boxes mark the interface states. Here $\beta-\beta_0$ is given in units of $10^{-4}k_0$, that is, as the effective-index difference $\neff-n_0$ in units of $10^{-4}$, where $n_0=\beta_0/k_0$ is the mean effective index of the interface states.}
\label{fig:1}
\end{figure*}

Replacing the central waveguide in Fig.~\ref{fig:1}(c) with a hollow waveguide introduces two local OAM orbitals. In the ideal model, both orbitals couple identically to the neighboring Gaussian waveguides [insets (i) and (iii) of Fig.~\ref{fig:1}(d)]. After subtracting the common onsite propagation constant, the array Hamiltonian is
\begin{equation}
H_{\mathrm{OAM}} = \begin{pmatrix}
\ddots & b & 0 & 0 & 0 & 0 & 0 & 0 \\
b & 0 & a & 0 & 0 & 0 & 0 & 0 \\
0 & a & 0 & \kappa_1 & \kappa_1 & 0 & 0 & 0 \\
0 & 0 & \kappa_1 & 0 & h & \kappa_2 & 0 & 0 \\
0 & 0 & \kappa_1 & h & 0 & \kappa_2 & 0 & 0 \\
0 & 0 & 0 & \kappa_2 & \kappa_2 & 0 & a' & 0 \\
0 & 0 & 0 & 0 & 0 & a' & 0 & b' \\
0 & 0 & 0 & 0 & 0 & 0 & b' & \ddots
\end{pmatrix}.
\label{eq:HOAM}
\end{equation}
Here $\kappa_1$ and $\kappa_2$ couple either central OAM orbital to its left and right neighbors, respectively, and all displayed couplings are real. We first take matched onsite propagation constants and $h=0$. We seek eigenstates of the full array whose central field occupies only one of the two OAM orbitals. In the ideal model, such states satisfy
\begin{equation}
H_{\mathrm{OAM}}|\Psi_\sigma\rangle = 0, \qquad \Pi_c|\Psi_\sigma\rangle = C|\sigma l\rangle, \qquad \sigma=\pm1.
\label{eq:zeromode}
\end{equation}
Here $\Pi_c$ projects onto the central waveguide and $C$ is the central amplitude of a normalized array eigenstate. The zero eigenvalue is measured relative to the common propagation constant $\beta_0$, so the two states are degenerate while their central fields carry opposite OAM charges. The definite charge refers to the central field; the surrounding waveguides carry evanescent tails, and the array as a whole is not rotationally symmetric. Because the two states share the same tails, they are not orthogonal; their overlap is $1-|C|^2$.

These eigenstates can be constructed explicitly. Label each arm outward from the interface as $B_1, A_1, \ldots, B_M, A_M$, with coupling $t_2$ between $B_j$ and $A_j$ and $t_1$ between $A_j$ and $B_{j+1}$. This outward labeling does not redefine the bulk unit cells; in the notation of Fig.~\ref{fig:1}(c), $t_2=a$ and $t_1=b$ on the left arm, and $t_2=a'$ and $t_1=b'$ on the right arm. Let $\kappa_s$ be the coupling from either central orbital to the first site on arm $s$, where $s=\mathrm{L},\mathrm{R}$. For either choice of the occupied central OAM orbital, the zero-mode equations are solved by
\begin{equation}
b_{s,j} = 0, \qquad a_{s,j} = -\frac{\kappa_s}{t_2}\left(-\frac{t_1}{t_2}\right)^{j-1} C.
\label{eq:tails}
\end{equation}
The amplitudes $a_{s,j}$ and $b_{s,j}$ belong to the $A_j$ and $B_j$ sites, respectively, and are distinct from the bulk couplings $a$ and $b$. The B sites carry no field. The zero-mode equation at the first B site fixes the first A amplitude by balancing the field from the occupied central orbital, and the remaining equations give the recurrence in Eq.~\eqref{eq:tails}. Consequently, for $|t_1/t_2|<1$, the field decays into both arms and the state is localized at the interface. We call this termination, in which the first bond of each arm is the stronger one, the S-interface, and the reversed termination the W-interface. Reversing the arm termination can move the finite-chain zero-mode weight toward the remote ends, so the zero-mode count alone does not guarantee localization at the central waveguide. The normalization and the S/W comparison are derived in Supplemental Material Secs.~S3.1 and S4~\cite{SM}.

Equations~\eqref{eq:zeromode} and \eqref{eq:tails} define the OAM interface eigenstates directly, independent of how a numerical solver chooses a basis within the degenerate subspace. Under propagation, each eigenstate acquires only the common phase $\exp(i\beta_0 z)$, so its spatial profile and central vortex phase are preserved; coupling to the neighboring Gaussian waveguides does not spread it into extended bulk modes.

The structure of these states is clearest in a bright--dark basis. Because the two OAM orbitals couple identically to the arms, only their symmetric combination $|B\rangle = (|{+l}\rangle + |{-l}\rangle)/\sqrt{2}$ couples to the array, with strength $\sqrt{2}\kappa_s$; the antisymmetric combination $|D\rangle = (|{+l}\rangle - |{-l}\rangle)/\sqrt{2}$ is decoupled and forms a zero mode confined to the hollow waveguide. The bright orbital acts as the central site of a conventional SSH domain wall and supports the topological interface mode, whereas the dark mode exists for any arm couplings. The OAM eigenstates in Eq.~\eqref{eq:zeromode} are superpositions of these two zero modes, so their definite central charge relies on the degeneracy of a topological mode and a symmetry-enforced dark mode, both fixed at zero by chiral symmetry. For $N$ central orbitals with matched propagation constants and identical coupling profiles, the same argument gives one bright and $N-1$ dark orbitals, and hence $N$ zero modes. This count equals the sublattice imbalance of the array and reflects the enlarged interface space rather than an increase in the bulk winding number (Ref.~\cite{Ramachandran2017} and Supplemental Material Sec.~S2.2~\cite{SM}); the extension to longer-range couplings is discussed in Supplemental Material Sec.~S2.3.

The ideal counts of one, two, and six zero modes correspond to Figs.~\ref{fig:1}(c)--\ref{fig:1}(e), respectively. The spectra in Figs.~\ref{fig:1}(f)--\ref{fig:1}(h) are waveguide-mode calculations, whose interface-state splittings reflect the physical modal detunings and symmetry-breaking couplings. For an opposite-charge pair, rotational symmetry gives degenerate reference modes in the isolated hollow waveguide. For $N>2$, the central orbitals necessarily include modes of different angular or radial order, whose propagation constants are not related by symmetry and must be matched by design; the spread of the six interface states in Fig.~\ref{fig:1}(h) reflects this. The $N$-orbital model therefore serves mainly as a mode-counting reference, and the remainder of this work focuses on the opposite-charge pair.

\subsection{OAM confinement in coupled waveguide environments}
\label{sec:confinement}

\begin{figure*}[t]
\includegraphics[width=\textwidth]{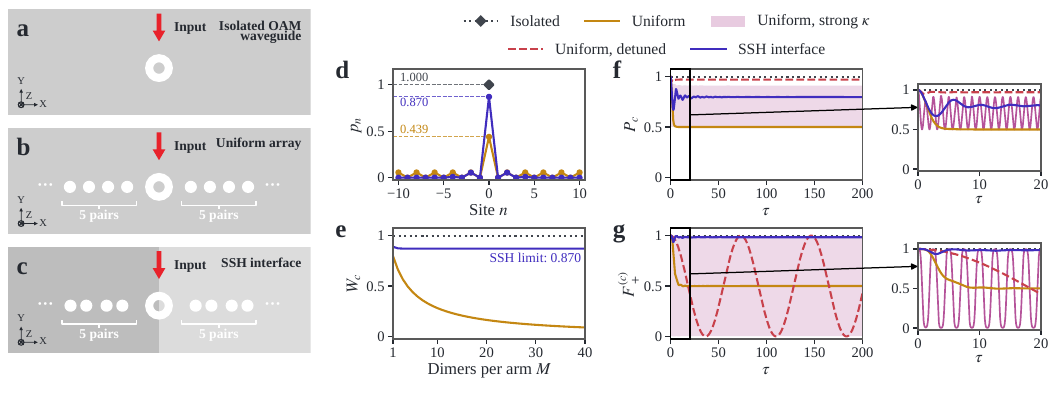}
\caption{OAM confinement and charge retention in isolated and coupled waveguide environments. (a)--(c) An isolated hollow waveguide, a hollow waveguide coupled to uniform Gaussian-waveguide arms, and an SSH interface. Ellipses abbreviate the arms; the schematics indicate five dimers per arm. Red arrows indicate central excitation. (d) Normalized eigenstate power $p_n$ per physical waveguide for $M=5$, choosing a zero-mode eigenstate with only charge $+l$ in the hollow waveguide. Dashed lines mark the central weights 1.000, 0.439, and 0.870. (e) Central eigenstate weight $W_c$ versus the number of dimers per arm; the SSH limit is given by Eq.~\eqref{eq:Wclimits}. (f) Central power $P_c$ and (g) conditional target fraction $\Fp$, defined in Eq.~\eqref{eq:observables}, for identical local $+l$ inputs with all Gaussian sites initially unexcited. Two further references use the uniform geometry of (b) with a stronger contact coupling, $\kappa=1.2J$ (shaded band: envelope of a fast oscillation), or with an onsite detuning $\delta=3J$ of the hollow waveguide (dashed line); insets resolve $0\le\tau\le20$. Neither reference has a localized zero mode with a single central charge, so both are absent from (d) and (e). Propagation curves use 250 dimers per arm; $\tau=Jz$. Unless stated otherwise, the coupled models use $\kappa=0.25J$, with $(t_2,t_1)=(J,0.4J)$ for SSH and $(0.7J,0.7J)$ for the uniform arrays. Total input power is unity. All calculations assume equal coupling profiles for opposite charges and no direct central OAM mixing.}
\label{fig:2}
\end{figure*}

An isolated hollow waveguide sets the reference for ideal OAM propagation. Embedding it in an array introduces additional coupling channels, and the question is how closely the hollow waveguide can retain the isolated-guide behavior while remaining coupled to the array. Detuning a waveguide from its neighbors also suppresses resonant power transfer; we include this case below as a further reference. We compare an isolated hollow waveguide, the same guide coupled to two uniform Gaussian-waveguide arms, and the SSH interface described above [Figs.~\ref{fig:2}(a)--\ref{fig:2}(c)], first through their eigenstates and then through propagation from the same local OAM input.

We use the lossless nearest-neighbor model of Eq.~\eqref{eq:HOAM} with matched onsite propagation constants, zero direct coupling between the central OAM orbitals, and equal real coupling $\kappa$ from either orbital to either arm. Each arm contains $M$ Gaussian dimers, ordered as in Eq.~\eqref{eq:tails} and terminated at the outermost A site. The isolated reference has $\kappa=0$. The coupled systems use $\kappa=0.25J$, with $t_2=J$ and $t_1=0.4J$ for the SSH interface and $t_2=t_1=t=0.7J$ for the uniform array, where $J$ is a coupling scale with units of inverse length. Both coupled systems thus have the same contact coupling and the same sum of bulk couplings. These parameters define a controlled model comparison and are not fitted to the waveguide geometry used in the subsequent simulations.

First, consider a normalized zero-mode eigenstate $|\Psi_+\rangle$ whose central component contains only charge $+l$. Equation~\eqref{eq:tails} fixes the Gaussian tails relative to the central amplitude $C$. Defining $q=t_1/t_2$ and summing the central power and the Gaussian-site powers over both arms gives the central eigenstate weight
\begin{equation}
W_c(M) = |C|^2 = \left[1 + \frac{2\kappa^2}{t_2^2}\sum_{j=0}^{M-1} q^{2j}\right]^{-1}.
\label{eq:Wc}
\end{equation}
For the SSH termination with $q<1$, the geometric series converges, whereas for the uniform arms $q=1$ every occupied Gaussian site has the same tail amplitude. Consequently,
\begin{equation}
\begin{aligned}
W_c^{\mathrm{SSH}}(\infty) &= \left[1 + \frac{2\kappa^2}{t_2^2 - t_1^2}\right]^{-1}, \\
W_c^{\mathrm{uniform}}(M) &= \left[1 + \frac{2M\kappa^2}{t^2}\right]^{-1}.
\end{aligned}
\label{eq:Wclimits}
\end{equation}
The SSH eigenstate therefore retains a finite central weight as the arms grow, whereas the uniform-array weight decreases to zero. For $M=5$, the central weights are 1.000, 0.439, and 0.870 for the isolated, uniform, and SSH systems, respectively [Fig.~\ref{fig:2}(d)], and the SSH value saturates at 0.870 within a few dimers [Fig.~\ref{fig:2}(e)]. Here $p_n$ denotes power in a physical waveguide, with the two central orbital powers added at $n=0$. The dark mode exists in both coupled systems, but it contains both charges equally and therefore does not by itself provide a definite central charge. In the uniform array, the bright zero mode is extended, so the central weight vanishes as $M\to\infty$.

Second, we launch the same local input in all systems: $c_+(0)=1$, with the opposite central OAM orbital and all Gaussian sites initially unexcited. Because this input lacks the tails in Eq.~\eqref{eq:tails}, it differs from the array eigenstate and redistributes initially even in the SSH system. We propagate with the evolution operator $\exp(iHz)$, consistent with Eq.~\eqref{eq:cmt}, and use the dimensionless distance $\tau=Jz$. We record the total power remaining in the central waveguide and the fraction of it carried by the target charge:
\begin{equation}
P_c(\tau) = |c_+(\tau)|^2 + |c_-(\tau)|^2, \quad \Fp(\tau) = |c_+(\tau)|^2/P_c(\tau).
\label{eq:observables}
\end{equation}
The first measure captures leakage into the arms; the second measures the central modal composition and thus conversion to the opposite charge. The propagation calculation uses 250 dimers per arm to exclude end reflections over $0\le\tau\le200$; increasing this number to 300 leaves the plotted curves unchanged at the displayed precision. The isolated reference remains at unity in both measures. At $\tau=60$, the SSH system retains $P_c=0.80$ and $\Fp=0.984$, compared with 0.50 and 0.50 in the uniform array [Figs.~\ref{fig:2}(f) and \ref{fig:2}(g)]. The oscillatory transient reflects excitation of additional array modes by the local input. The transient is bounded: for these parameters the target-charge fraction stays above 0.91 and the central power above 0.64 at every distance (Supplemental Material Sec.~S3.6~\cite{SM}).

These values follow from the bright--dark decomposition. The input $|{+l}\rangle = (|B\rangle + |D\rangle)/\sqrt{2}$ places half of its power in the dark mode, which stays in the hollow waveguide in both coupled systems. In the SSH array, the bright half projects onto the bright zero mode, whose central weight is $W_b = [1 + 4\kappa^2/(t_2^2 - t_1^2)]^{-1}$, and the remainder disperses into the arms. This weight is related to the central weight of the eigenstate with a definite central charge by $W_c = 2W_b/(1+W_b)$, or $W_b = W_c/(2-W_c)$ (Supplemental Material Sec.~S5.2~\cite{SM}). Here $W_c$ characterizes the eigenstate with a definite central OAM charge, whereas $W_b$ is the central weight of the normalized bright zero mode; the value 0.870 thus refers to the stationary eigenstate, and the central power of about 0.80 to the evolution of a local OAM input. In the long-arm limit, after the dispersive contribution has dephased locally and before boundary recurrences, the central amplitudes approach $|c_\pm| = (1\pm W_b)/2$, giving $P_c\to(1+W_b^2)/2$ and $\Fp\to(1+W_b)^2/[2(1+W_b^2)]$. For the SSH parameters, $W_b=0.771$, so that $P_c\to0.797$ and $\Fp\to0.984$, which Figs.~\ref{fig:2}(f) and \ref{fig:2}(g) approach. In the uniform array, $W_b\to0$, and both measures reduce to 1/2, the contribution of the dark mode. The SSH interface thus improves OAM retention by localizing the bright component; because the bright and dark components remain degenerate, they keep their relative phase, and conversion to the opposite charge is suppressed.

The role of bright--dark degeneracy can be separated from bright-mode localization using two additional references. Both references support localized bright components [Figs.~\ref{fig:2}(f) and \ref{fig:2}(g); Table~\ref{tab:references}]. With a strong contact coupling, $\kappa=1.2J$, the bright combination forms bound states outside the arm band at $E=\pm2.51J$; the model remains chiral, but these states are not degenerate with the dark mode. With the original coupling but an onsite detuning $\delta=3J$ of the hollow waveguide, chiral symmetry is broken and the bright bound state is shifted from the dark mode by $\Sigma=0.086J$. The strong-contact and detuned references have time-averaged $P_c=0.71$ and 0.97, respectively, compared with 0.80 for SSH. Thus the detuned reference retains more central power than SSH, while both references undergo nearly complete conversion between $+l$ and $-l$, with conversion lengths $\pi/E=1.25/J$ and $\pi/\Sigma=36.7/J$, respectively [Fig.~\ref{fig:2}(g)]. For the detuned reference, $\Sigma\approx4\kappa^2/\delta$, so that $\LB\approx\pi\delta/(4\kappa^2)$ (Supplemental Material Sec.~S5.4~\cite{SM}): increasing the detuning reduces hybridization with the arms and lengthens the charge-conversion scale, but only linearly, so that retaining the charge over a length $L$ requires $\delta\gg4\kappa^2L/\pi$, which amounts to decoupling the hollow waveguide from the array. The cost of this route is clearest at equal conversion length. A residual mismatch $\delta_c$ between the hollow and Gaussian waveguides splits the SSH doublet by $(1-W_b)\delta_c$, while the bright interface mode keeps a fraction $1-W_b\approx0.23$ of its power on the arms. A detuned waveguide with the same $\LB$ keeps only $(\pi/2\kappa\LB)^2$ on the arms, about $8\times10^{-5}$ for $\LB=685/J$, the conversion length produced by $\delta_c=0.02J$ (Supplemental Material Sec.~S5.4). In all of these models the two charges couple identically to the arms, and a localized bright component at the dark-mode propagation constant is then both necessary and sufficient for a stationary superposition with definite central charge. In the SSH interface, chiral symmetry fixes the bright interface mode at zero and the bulk gap localizes it, so both conditions are met while the hollow waveguide remains coupled to the array.

\begin{table*}[t]
\caption{Localization and charge retention for the local $+l$ input of Fig.~\ref{fig:2}. $\langle P_c\rangle$ is averaged over $100\le\tau\le200$; $\Fp$ is the value, or the range of oscillation, over the same interval; $\LB$ is the characteristic distance for conversion from $+l$ to $-l$.}
\label{tab:references}
\begin{ruledtabular}
\begin{tabular}{lllllll}
System & Chiral symmetry & Bright mode localized & Degenerate with dark mode & $\langle P_c\rangle$ & $\Fp$ & $\LB J$ \\
\colrule
Isolated & --- & --- & --- & 1 & 1 & $\infty$ \\
Uniform & Yes & No (extended) & Yes (extended) & 0.50 & 0.50 & --- \\
Uniform, $\kappa=1.2J$ & Yes & Yes ($\pm2.51J$) & No & 0.71 & 0--1 & 1.25 \\
Uniform, $\delta=3J$ & No & Yes ($\delta+0.086J$) & No & 0.97 & 0--1 & 36.7 \\
SSH interface & Yes & Yes (0) & Yes & 0.80 & 0.98 & $\infty$ (ideal) \\
\end{tabular}
\end{ruledtabular}
\end{table*}

The zero-mode construction also describes arrays with bond disorder. For the finite-chain connectivity used here, nonzero $t_2$ couplings keep the off-diagonal block at full column rank, fixing the zero-mode count to the sublattice imbalance of two. This holds for independently varied arm and contact couplings, including unequal contacts to the two OAM charges. The two zero modes then remain exactly degenerate, and a zero mode with a single occupied central charge still exists, because the zero-mode equations fix the arm amplitudes uniquely from the two central amplitudes. With independent contact disorder, a compact central dark state need not remain; the coupled-site recurrence supplies the pure-central-charge zero modes directly. In the model of Fig.~\ref{fig:2}, over 200 realizations with $\pm20\%$ random variation in all arm couplings and $\pm30\%$ variation in each contact coupling, the doublet stays degenerate to numerical precision and $\Fp$ at $\tau=1000$ averages 0.98 (minimum 0.90; Supplemental Material Sec.~S6~\cite{SM}). The three chiral-symmetry-breaking perturbations examined here---central onsite detuning, direct OAM mixing $h$, and coupling from the hollow waveguide to the first A sites of the arms---lift the degeneracy and induce OAM beating with characteristic conversion length $\LB=\pi/\Delta\beta$, where $\Delta\beta$ is the perturbed doublet splitting. Spatial confinement thus follows from the SSH termination, whereas suppressing conversion in this doublet-beating regime over a length $L$ additionally requires $L\ll\LB$, which depends on the modal symmetry and propagation-constant matching. Direct OAM mixing is not suppressed by the interface: by Eq.~(S38) of the Supplemental Material it splits the doublet by $(1+W_b)h\approx1.77h$, close to the splitting $2h$ of an isolated hollow waveguide. The protection discussed here therefore covers bond disorder and confinement, not mixing within the doublet.

\section{Simulations}
\label{sec:simulations}

\subsection{Numerical characterization of OAM interface modes}
\label{sec:modes}

\begin{figure*}[t]
\includegraphics[width=\textwidth]{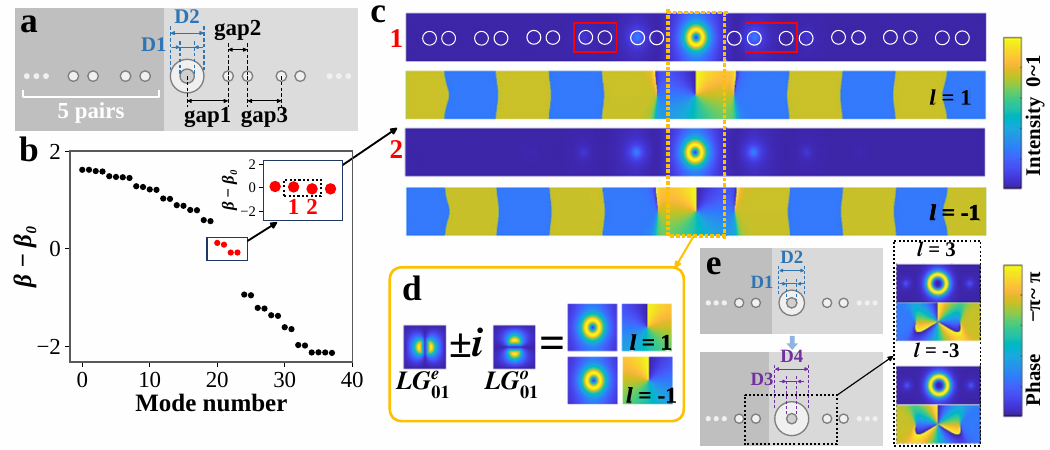}
\caption{Interface eigenmodes and OAM superpositions in the waveguide array. (a) S-interface geometry with five Gaussian-waveguide pairs on each side of a hollow waveguide. (b) Calculated propagation-constant spectrum, with $\beta-\beta_0$ in units of $10^{-4}k_0$ as in Fig.~\ref{fig:1}; the inset highlights four interface modes, two per polarization. Labels 1 and 2 identify the selected $x$-polarized doublet. (c) Intensity and phase profiles of the OAM superpositions with $l=+1$ and $l=-1$. (d) Construction of opposite-charge OAM fields from even and odd standing-wave components with a relative phase of $\pm\pi/2$. (e) A larger hollow-waveguide geometry and the corresponding higher-order OAM fields with $l=\pm3$.}
\label{fig:3}
\end{figure*}

We next examine the physical waveguide modes of the S-interface geometry in Fig.~\ref{fig:3}(a), which has five pairs of Gaussian waveguides on each side of a central hollow waveguide with inner and outer diameters $D_1=12$\um\ and $D_2=20$\um. All waveguides have a core index of 1.442 in a background of index 1.440. The Gaussian-waveguide radius, 3.55\um, is chosen so that the effective index of its fundamental mode matches that of the OAM modes in the hollow waveguide, which implements the propagation-constant matching of the model. Modes are calculated with a full-vectorial finite-difference eigensolver (Lumerical MODE) at a wavelength of 1.55\um, and propagation is simulated with Lumerical FDTD. The doublet splitting and the propagation are cross-checked with an independent full-vectorial mode solver and a beam-propagation calculation (Supplemental Material Sec.~S7~\cite{SM}). The interface modes are quasi-linearly polarized (see below), so we describe them in the scalar approximation, in which the vector modes of the hollow waveguide combine into $x$- and $y$-polarized OAM pairs. The physical-array Hamiltonian is described in Supplemental Material Sec.~S2.4, and the dependence on chain length is discussed in Supplemental Material Sec.~S3.

The three separations are labeled $\mathrm{gap}_1$, $\mathrm{gap}_2$, and $\mathrm{gap}_3$ in Fig.~\ref{fig:3}(a). Here $\mathrm{gap}_1$ is the separation between the hollow waveguide and its nearest Gaussian neighbor, and the other two alternate along the Gaussian-waveguide arms. Measured edge to edge, the separation is 16\um\ between the hollow waveguide and its nearest Gaussian neighbor and between adjacent pairs, and 8\um\ within a pair. The S-interface has $\mathrm{gap}_1,\mathrm{gap}_3>\mathrm{gap}_2$. The comparison geometry with $\mathrm{gap}_1,\mathrm{gap}_3<\mathrm{gap}_2$ is denoted the W-interface; its localization and its stronger OAM mixing are discussed in Supplemental Material Sec.~S4.

The calculated spectrum in Fig.~\ref{fig:3}(b) contains four interface modes within the bulk gap, two for each polarization. We examine the two $x$-polarized modes labeled 1 and 2, whose TE polarization fractions of 100\% and waveguide TE/TM fractions above 99.9\% confirm quasi-linear polarization. The eigensolver returns standing-wave (even and odd) profiles; in the degenerate limit, any combination of them is an eigenstate, including the OAM states of Eq.~\eqref{eq:zeromode}. Consistent with the bright--dark picture, one mode of the doublet is confined to the hollow waveguide (effective area 543\,$\mu$m$^2$), whereas the other extends into the arms (1056\,$\mu$m$^2$). In the physical array, residual detuning and OAM mixing split each doublet. The splitting in the original MODE spectrum, 6--$7\times10^{-6}$, is sensitive to the computational domain and mesh; refined calculations with an independent full-vectorial solver give about $1.1\times10^{-5}$ for the $x$-polarized doublet and $1.2\times10^{-5}$ for the $y$-polarized doublet on the finest mesh, corresponding to a characteristic OAM conversion length $\LB=\lambda/(2\Delta\neff)\approx65$--70~mm, about four times the 15-mm propagation length considered below (Supplemental Material Secs.~S3.2, S3.5, and S7~\cite{SM}). Part of this splitting comes from the array itself: the Gaussian waveguides lie along the array axis and therefore perturb the two standing-wave orbitals of the hollow waveguide differently, and in the OAM basis this difference acts as a direct mixing. Because central detuning and mixing both enter the splitting linearly, a small detuning of the Gaussian waveguides can compensate this term (Supplemental Material Secs.~S4 and S7).

For field reconstruction and source preparation, the central even and odd profiles are represented by real Laguerre--Gaussian-type components $LG^{\mathrm{e}}_{0l}$ and $LG^{\mathrm{o}}_{0l}$~\cite{Kimel1993}. Combining them with a relative phase of $\pm\pi/2$ and amplitudes matched to their local weights gives the opposite-charge OAM fields [Fig.~\ref{fig:3}(d)]. This is a change of basis within the doublet, not an additional localization mechanism. Figure~\ref{fig:3}(c) shows the resulting $l=\pm1$ intensity and phase profiles; the annular intensity and azimuthal phase winding identify the central vortex fields.

A larger hollow waveguide, with inner and outer diameters $D_3>D_1$ and $D_4>D_2$, supports higher-order OAM fields with charges $l=\pm3$ [Fig.~\ref{fig:3}(e)]; the propagation analysis below is restricted to the lower-order pair.

\subsection{Propagation of opposite-charge OAM fields}
\label{sec:propagation}

\begin{figure}[t]
\includegraphics[width=\columnwidth]{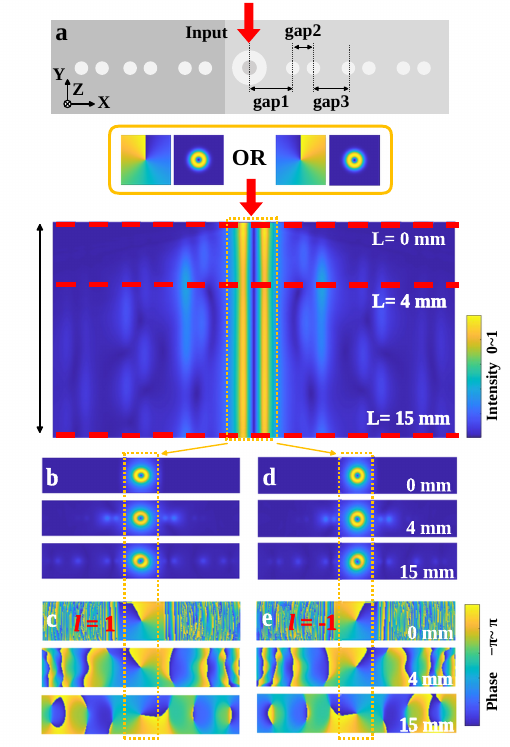}
\caption{Propagation of OAM fields through the S-interface array. (a) Excitation scheme and simulated longitudinal intensity distribution in the $xz$ plane over 15~mm. (b), (c) Transverse intensity and phase profiles for an $l=+1$ input. (d), (e) Corresponding profiles for an $l=-1$ input. The transverse planes are located at $z=0$, 4, and 15~mm.}
\label{fig:4}
\end{figure}

We next simulate propagation in the physical S-interface array over 15~mm. Fields with $l=+1$ and $l=-1$ are launched separately into the central hollow waveguide, as illustrated in Fig.~\ref{fig:4}(a). Each input is a superposition of the two $y$-polarized first-order modes of the hollow waveguide with a relative phase of $\pm90^\circ$, so Fig.~\ref{fig:4} excites the $y$-polarized doublet; the simulation settings are given in Supplemental Material Sec.~S3.2~\cite{SM}. The longitudinal intensity map shows initial spreading into neighboring waveguides, followed by confinement near the interface over the full simulated distance. As in the model of Fig.~\ref{fig:2}, the source occupies only the central waveguide and therefore also excites modes other than the interface doublet, which accounts for the initial redistribution.

Figures~\ref{fig:4}(b)--\ref{fig:4}(e) show transverse intensity and phase profiles at $z=0$, 4, and 15~mm for the two inputs. At each plane, the central field retains an annular intensity distribution and the phase winding of the input charge, consistent with the OAM interface eigenstates of the ideal model. An independent scalar beam-propagation calculation, which treats the same geometry in the weak-guidance approximation and does not resolve polarization, quantifies this retention: at 15~mm, the launched charge carries 86\% of the power within 16\um\ of the axis, angular orders other than $l=\pm1$ carry less than $10^{-4}$, and the conversion follows a characteristic length of 66~mm, consistent with the finest-grid mode splitting (Supplemental Material Sec.~S7~\cite{SM}). Because the launched charge keeps the larger weight, the phase winding keeps the sign of the input over the simulated length.

\section{Conclusion}
\label{sec:conclusion}

A hollow waveguide embedded at an SSH domain wall provides a multimode interface for localized OAM transport. In the ideal model, the interface supports two degenerate eigenstates whose fields in the hollow waveguide carry opposite OAM charges, each a superposition of the topological interface mode and a dark mode of the hollow waveguide. The comparison with uniform, strongly coupled, and detuned arrays separates two requirements: the SSH gap localizes the bright component, and chiral symmetry keeps it degenerate with the dark mode, so that a launched charge is both confined and retained while the hollow waveguide remains phase-matched to the array. Mode and propagation simulations of a physical array show confined $l=\pm1$ vortex fields that keep their phase winding over 15~mm.

The degeneracy is exact under bond disorder, whereas central detuning, direct OAM mixing, and same-sublattice coupling split the doublet; the splitting $\Delta\beta$ then sets the conversion length $\LB=\pi/\Delta\beta$. In the simulated array, the finest-grid splittings of 1.1--$1.2\times10^{-5}$ for the two polarizations give $\LB\approx65$--70~mm, and an independent propagation calculation gives about 14\% conversion over 15~mm. For more than two central orbitals, the interface hosts $N$ zero modes provided that $N$ modes of different order are matched by design. Because central detuning and OAM mixing enter the splitting linearly, the geometry can be tuned so that they cancel, which lengthens $\LB$ (Supplemental Material Sec.~S7~\cite{SM}). The construction thus provides a tunable design route for integrating OAM-supporting waveguides into coupled photonic structures.\par\vspace{0pt plus 1fill}


\bibliography{refs}

\end{document}


\title{Supplemental Material for ``On-chip topological orbital angular momentum modes at a multimode interface''}
\author{Xuyang Li}
\affiliation{Department of Materials Science and Engineering, University of Pennsylvania, Philadelphia, Pennsylvania 19104, USA}
\maketitle

This supplement develops the coupled-mode model of a hollow-waveguide interface between SSH chains, constructs its OAM-carrying eigenstates, and provides the analytical and numerical methods supporting main-text Figs.~1--4 and Tables~I and II.

The ideal Hamiltonian calculations are distinguished from the physical waveguide-mode and propagation simulations. Sections~\ref{sec:S2}--\ref{sec:S4} specify the modal symmetry, termination, and matching assumptions. Section~\ref{sec:S5} derives and documents the comparison in main-text Fig.~2 and Table~II, including the strongly coupled and detuned references, and Sec.~\ref{sec:S6} examines robustness to chiral-preserving and chiral-breaking perturbations. Figures~S1--S5 show calculations in the normalized coupled-mode model, and Sec.~\ref{sec:S7} reports independent checks of the physical-array calculations.

\tableofcontents

\section{Comparison of ordinary and multimode interfaces}
\label{sec:S1}

The comparison in main-text Table~I concerns the local mode spaces used to describe the lattice. In a conventional single-orbital SSH description, each bulk site and the interface site carry the same type of optical mode. A domain wall with a unit winding-number difference supports a corresponding interface mode under the usual gap, symmetry and termination assumptions. Here, the Gaussian bulk waveguides retain one relevant spatial mode per polarization, whereas the hollow interface waveguide supplies an opposite-charge OAM pair.

This distinction can be expressed as a change in the local Hilbert space at the junction while leaving the asymptotic bulk chains unchanged. For the physical example with charges $l=\pm1$, the local interface space has dimension two per polarization. The $N$-orbital construction in main-text Fig.~1(e) is an abstract extension of this interface space; it does not establish that arbitrarily many OAM orders in one hollow waveguide have identical propagation constants.

The comparison in main-text Table~I is made at the level of the local mode basis; it does not imply that every one-dimensional topological system has only one state. The following cited cases illustrate the distinction.

In the waveguide-array example of main-text Ref.~[38], two topological interfaces are coupled, allowing oscillations between interface states~\cite{SR1}. Its multiplicity therefore includes spatially distinct interfaces, whereas the present construction enlarges the orbital space at a single junction.

For the plasmonic example of main-text Ref.~[21], kinked and antikinked SSH interfaces can both support interface eigenmodes, but a single-waveguide input excites them differently~\cite{SR2}. Thus excitation accessibility must be distinguished from existence of an eigenstate. The same distinction matters when interpreting the OAM propagation figure here.

For the nanocavity example of main-text Ref.~[22], identical L3 cavities with alternating couplings provide the local resonator basis, and finite terminations determine the edge states~\cite{SR3}. The present interface instead supplies an OAM doublet while the outer sites retain a Gaussian-mode description.

The table compares the local mode basis at the boundary of selected one-dimensional topological arrays; it does not count all localized states of the finite structures. For example, the antikink structure of main-text Ref.~[21] supports three zero modes, one at the interface and two at the ends.

\section{Coupled-mode theory and interface Hamiltonians}
\label{sec:S2}

\subsection{Coupling between Gaussian and OAM modes}
\label{sec:S21}

For a fixed polarization in a lossless, longitudinally invariant, weakly guiding medium, the scalar field satisfies the Helmholtz equation. We use the longitudinal phase convention $\exp(i\beta z)$, consistent with main-text Eq.~(1). The scalar approximation neglects polarization coupling; the physical modes in Sec.~\ref{sec:S32} are computed with a full-vectorial solver.
\begin{equation}
\left(\nabla_\perp^2 + \partial_z^2 + k_0^2 n^2\right)E = 0, \qquad k_0 = 2\pi/\lambda.
\end{equation}

Let $f_\mu(x,y)$ be a localized basis mode and expand the field as $E(x,y,z) = \exp(i\beta_0 z)\sum_\mu a_\mu(z) f_\mu(x,y)$. Neglecting the second derivative of the slowly varying amplitudes and projecting onto the transverse basis gives a generalized coupled-mode equation. The overlap matrix is $S_{\nu\mu} = \int f_\nu^* f_\mu\,dA$ and the projected transverse operator is $K_{\nu\mu} = (2\beta_0)^{-1}\int f_\nu^*\left[\nabla_\perp^2 + k_0^2 n^2 - \beta_0^2\right] f_\mu\,dA$.

For completeness, define the transverse operator $L_\perp = \nabla_\perp^2 + k_0^2 n^2$. Substitution into the Helmholtz equation gives the following expression before the paraxial reduction; primes denote derivatives with respect to $z$.
\begin{equation}
\sum_\mu \left[a_\mu'' + 2i\beta_0 a_\mu' + \left(L_\perp - \beta_0^2\right)a_\mu\right] f_\mu = 0.
\end{equation}

Dropping $a_\mu''$ and multiplying by $f_\nu^*$ followed by transverse integration yields the projected equation below. The approximation requires longitudinal envelope variation slow compared with $\beta_0$, and a mode basis adequate for the considered fields.
\begin{equation}
-i\,S\,\partial_z a = K a.
\end{equation}

For a linearly independent nonorthogonal basis, introduce $b = S^{1/2}a$. Then $-i\partial_z b = Hb$ with $H = S^{-1/2}KS^{-1/2}$. This orthonormalized Hamiltonian is Hermitian for the lossless scalar model. In the weak-overlap limit $S$ is approximately the identity, yielding the usual coupled-mode form. The diagonal entries are detunings from $\beta_0$; restoring the common phase recovers the propagation-constant convention of the main text.

For a $z$-independent basis, $K = K\Tr$ and $S = S\Tr > 0$ imply conservation of $a\Tr S a$, since the two terms in its $z$ derivative cancel. This is the modal norm of the scalar theory. If physical electromagnetic power is required, the vector-mode normalization must be used; a scalar norm is not automatically the exact Poynting flux.
\begin{equation}
\partial_z\left(a\Tr S a\right) = -i\,a\Tr K a + i\,a\Tr K a = 0.
\end{equation}

For a perturbation $\delta n^2$ of an isolated-guide reference operator, the corresponding first-order coupling in this normalization is $\kappa_{\nu\mu} \simeq [k_0^2/(2\beta_0)]\int f_\nu^*\,\delta n^2 f_\mu\,dA$. A vector calculation instead requires the appropriate power-normalized electromagnetic overlap.

Within an approximately circular hollow guide, write $f_{\pm l}(r,\varphi) = R_l(r)\exp(\pm il\varphi)$. In the basis $(f_{\mathrm{G}}, f_{+l}, f_{-l})$, the projected Hamiltonian contains the three onsite detunings $\delta_{\mathrm{G}}$, $\delta_+$, $\delta_-$, Gaussian-to-OAM couplings $g_+$, $g_-$, and an intra-doublet mixing $h$. Hermiticity requires reverse couplings to be complex conjugates. Equality of the two Gaussian-to-OAM couplings additionally depends on geometry and basis phases; reciprocity alone does not set all couplings to the same real number.
\begin{equation}
H = \begin{pmatrix}
\delta_{\mathrm{G}} & g_+ & g_- \\
g_+^* & \delta_+ & h \\
g_-^* & h^* & \delta_-
\end{pmatrix}.
\end{equation}

Under a mirror reflection $\varphi\to-\varphi$, the scalar OAM functions are exchanged. A Gaussian reference mode on the reflection axis can be chosen even. For a reflection-invariant coupling operator, applying this symmetry to the overlap integral equates $g_+$ and $g_-$ on that axis. Hermiticity then fixes the reverse matrix elements. This provides the explicit symmetry step required in addition to reciprocity.

For a rotationally related Gaussian guide located at azimuth $\theta$, and with $\kappa$ defined as $\langle G_\theta|H|{\pm l}\rangle$, a compatible local phase convention gives $g_\pm(\theta) = g_0\exp(\pm il\theta)$. Along the $x$ axis, $\theta = 0$ or $\pi$, so both charges acquire the same factor on each side, with $\exp(\pm il\pi) = (-1)^l$. Left and right coupling vectors are then parallel in the ideal symmetric geometry, allowing the equal-profile interface model after suitable site rephasing.

For two neighboring Gaussian guides at angles $\theta_{\mathrm{L}}$ and $\theta_{\mathrm{R}}$, the matrix $V$ mapping the central OAM amplitudes to the two neighbors can be written as follows, assuming both radial overlaps $g_s$ are nonzero.
\begin{gather}
V = \begin{pmatrix}
g_{\mathrm{L}}\exp(il\theta_{\mathrm{L}}) & g_{\mathrm{L}}\exp(-il\theta_{\mathrm{L}}) \\
g_{\mathrm{R}}\exp(il\theta_{\mathrm{R}}) & g_{\mathrm{R}}\exp(-il\theta_{\mathrm{R}})
\end{pmatrix}, \\
\det V = 2i\,g_{\mathrm{L}}g_{\mathrm{R}}\sin\!\left[l(\theta_{\mathrm{L}} - \theta_{\mathrm{R}})\right].
\end{gather}
Hence a compact central dark combination exists when $l(\theta_{\mathrm{L}} - \theta_{\mathrm{R}})$ is an integer multiple of $\pi$. A collinear placement satisfies this condition for integer $l$. A generic angular displacement can give $\operatorname{rank}V = 2$ and remove the compact dark combination, while finite-array zero modes can still exist by sublattice imbalance. Loss of compactness and loss of zero modes are therefore different statements.

\subsection{Finite SSH chains and the interface-orbital construction}
\label{sec:S22}

For a bulk chain with intracell and intercell couplings $a$ and $b$ (main-text notation), the off-diagonal Bloch element is $\eta(k) = a + b\exp(ik)$. The chiral operator $\Gamma = \mathrm{diag}(1,-1)$ anticommutes with the Bloch Hamiltonian. In a fixed unit-cell convention, the winding number counts the winding of $\eta(k)$ around the origin and changes at $|a| = |b|$. Comparison of the two domains requires a consistent real-space termination and sublattice convention.
\begin{gather}
H(k) = \begin{pmatrix} 0 & \eta^*(k) \\ \eta(k) & 0 \end{pmatrix}, \\
\varepsilon_\pm(k) = \pm\sqrt{a^2 + b^2 + 2ab\cos k}.
\end{gather}
For positive real couplings, $\eta(k)$ traces a circle of radius $b$ centered at $a$. It encloses the origin when $b > a$ and does not when $b < a$, giving winding numbers one and zero in this convention. The full bulk gap is $2|b-a|$. These statements are bulk properties; assigning an interface state requires the specified junction and terminations.
\begin{equation}
w = \frac{1}{2\pi}\int_{-\pi}^{\pi}\partial_k\arg\eta(k)\,dk.
\end{equation}

An explicit finite model makes the state count and boundary conditions unambiguous. Label the $N$ central orbitals $c_\mu$ and the sites in arm $s = \mathrm{L},\mathrm{R}$, moving outward from the interface, as $B_{s,1}, A_{s,1}, \ldots, B_{s,M}, A_{s,M}$. Assign $c_\mu$ and $A_{s,j}$ to sublattice A and $B_{s,j}$ to sublattice B. Within each arm, couple $B_{s,j}$ to $A_{s,j}$ by $t_2$ and $A_{s,j}$ to $B_{s,j+1}$ by $t_1$. Couple every $c_\mu$ to $B_{s,1}$ by $\kappa_s$. In the notation of main-text Fig.~1(c), $t_2 = a$ and $t_1 = b$ on the left arm, and $t_2 = a'$ and $t_1 = b'$ on the right arm; the outward labels do not redefine the bulk unit cells. All onsite detunings and couplings among $c_\mu$ are set to zero in the ideal model. Its dimension is $N + 4M$; $M = 5$ gives 21 sites for $N = 1$ and 22 for $N = 2$.

The following compact operator expression specifies all nonzero entries of the ideal Hermitian matrix; H.c.\ denotes the Hermitian conjugate of the preceding hopping terms.
\begin{align}
H_{\mathrm{arm}} &= \sum_{s,j} t_2\,|B_{s,j}\rangle\langle A_{s,j}| + \mathrm{H.c.}, \\
H_{\mathrm{link}} &= \sum_s \sum_{j=1}^{M-1} t_1\,|A_{s,j}\rangle\langle B_{s,j+1}| + \mathrm{H.c.}, \\
H_{\mathrm{int}} &= \sum_s \sum_{\mu=1}^{N} \kappa_s\,|B_{s,1}\rangle\langle c_\mu| + \mathrm{H.c.}
\end{align}

The symmetric central combination $|B\rangle = N^{-1/2}\sum_\mu|c_\mu\rangle$ couples to arm $s$ with amplitude $\sqrt{N}\kappa_s$. Every normalized central combination $|D\rangle = \sum_\mu d_\mu|c_\mu\rangle$ with $\sum_\mu d_\mu = 0$ is dark: all couplings to the arms cancel. There are $N-1$ independent dark zero modes. The bright central orbital is not itself a full-array eigenmode; the corresponding interface eigenmode includes field amplitudes on the neighboring A sites that decay outward when $|t_1/t_2| < 1$. These transverse field components are referred to as tails; they form part of the stationary eigenstate and do not represent continuing leakage during propagation.

An explicit orthonormal dark basis can be chosen with $\alpha = 1,\ldots,N-1$. Its coefficient sum vanishes, its norm is one, and the scalar product of two different members vanishes because the positive and negative contributions cancel.
\begin{equation}
|D_\alpha\rangle = \frac{\sum_{\mu=1}^{\alpha}|c_\mu\rangle - \alpha|c_{\alpha+1}\rangle}{\sqrt{\alpha(\alpha+1)}}.
\end{equation}

In the basis consisting of the bright orbital, all arm sites and the dark combinations, the Hamiltonian is the direct sum of a single-central-orbital chain with couplings $\sqrt{N}\kappa_s$ and an $(N-1)$-dimensional zero matrix. This proves both the compact dark-state count and the effective bright-chain description.

For real couplings in this outward arm convention, the bright zero-mode solution follows directly from the zero-energy site equations. The zero-energy equation on $B_{s,1}$ is $\sqrt{N}\kappa_s C + t_2 A_{s,1} = 0$. On $B_{s,j}$ with $j > 1$ it is $t_1 A_{s,j-1} + t_2 A_{s,j} = 0$. Setting every $B_{s,j}$ to zero satisfies the remaining central and A-site equations, including the outer termination. Iterating the B-site equations gives the solution below.
\begin{equation}
A_{s,j} = -\frac{\sqrt{N}\kappa_s}{t_2}\left(-\frac{t_1}{t_2}\right)^{j-1}C, \qquad B_{s,j} = 0.
\end{equation}
Here $C$ is the amplitude on the normalized bright central orbital. For $|t_1/t_2| < 1$, the mode decays outward from the junction. Its normalized central weight is
\begin{equation}
|C|^2 = \left[1 + N\sum_s\left|\frac{\kappa_s}{t_2}\right|^2\sum_{j=0}^{M-1}\left|\frac{t_1}{t_2}\right|^{2j}\right]^{-1}.
\label{eq:S16}
\end{equation}

A general finite-matrix proof follows by writing a chiral Hamiltonian as $H_1 = \begin{pmatrix} 0 & Q \\ Q\Tr & 0 \end{pmatrix}$, with $Q$ of size $n_A\times n_B$. Nonzero singular values of $Q$ generate paired positive and negative eigenvalues. Therefore the zero-mode dimension is $n_A + n_B - 2\operatorname{rank}Q$. Replacing one A-sublattice orbital by $N$ orbitals with exactly duplicated coupling rows leaves $\operatorname{rank}Q$ unchanged and adds $N-1$ to $n_A$.
\begin{align}
n_0(H_1) &= \dim\ker Q\Tr + \dim\ker Q, \\
n_0(H_1) &= n_A + n_B - 2\operatorname{rank}Q.
\end{align}
The two terms count zero modes residing on sublattices A and B. Duplicating a row neither adds an independent row nor removes the original row span. Thus $\operatorname{rank}Q_N = \operatorname{rank}Q_1$ and $n_A$ changes to $n_A + N - 1$, giving the identity below.
\begin{equation}
n_0(H_N) = n_0(H_1) + N - 1.
\label{eq:S19}
\end{equation}
For an original finite chain with exactly one zero mode, this identity gives $N$ zero modes and supports the $N = 1$, 2 and 6 ideal examples. The additional $N-1$ states have the local dark-state origin described above. The count includes all zero modes of the finite matrix; assigning them to a particular interface requires inspecting their spatial profiles.

For the arm termination used here, $n_A = 2M + N$ and $n_B = 2M$. Selecting the $A_{s,j}$ rows of $Q$ gives two triangular $M\times M$ blocks with diagonal $t_2$. If $t_2 \neq 0$, this submatrix is invertible and $Q$ has full column rank $2M$. Consequently the total finite-chain zero-mode count is exactly $N$ for any finite $M$. This does not by itself ensure that all $N$ modes are localized near the junction.

The $N = 1$, 2 and 6 normalized models were independently diagonalized with $M = 5$, $t_1/t_2 = 0.4$ and $\kappa_{\mathrm{L}}/t_2 = \kappa_{\mathrm{R}}/t_2 = 0.25$. The respective zero counts are 1, 2 and 6. The largest analytical-vector residual among these checks is $1.5\times10^{-16}$ in units $t_2$, and, for $N = 2$, the nearest nonzero eigenvalue has magnitude $0.6703857\,t_2$.

The $N$-orbital state-counting statement in main-text Sec.~II~C follows from Eq.~\eqref{eq:S19}. The pure-central-charge eigenstates in main-text Eq.~(4) are constructed separately by the amplitude cancellation described in Sec.~\ref{sec:S52} below. State counting alone does not establish their spatial localization or central OAM content.

\subsection{Scope of the extension to long-range hopping}
\label{sec:S23}

The finite-matrix rank argument does not require nearest-neighbor hopping. It remains valid if every hopping joins opposite sublattices and the complete coupling row of the replaced orbital, including long-range entries, is duplicated. These assumptions retain the block-off-diagonal form and its unchanged rank. They are stronger than allowing arbitrary long-range hopping.

Same-sublattice hopping, unequal onsite propagation constants and direct mixing between central orbitals generally invalidate the stated chiral form. Equal complex coupling profiles must also be distinguished from equal coupling magnitudes. For an infinite chain, a finite-matrix nullity argument alone does not prove spatial localization; a bulk gap, normalizability and suitable decay of the couplings must be checked.

More generally, if $V$ maps the $N$-dimensional central orbital space into the surrounding sites, and the central onsite block is zero, compact central zero modes belong to $\ker V$. Their number is $N - \operatorname{rank}V$. Rank-one coupling produces $N-1$ such modes. This statement counts the compact central states only and does not count every possible interface or remote-edge state of the complete array.

The reason the compact-state formula needs the zero central block is explicit: for a trial state $(c,0)$, the full eigenproblem gives $H_c c = \varepsilon c$ and $Vc = 0$. If $H_c = 0$, $\ker V$ supplies zero modes. If $H_c$ is nonzero, a vector in $\ker V$ is an eigenstate only when it also satisfies the central eigenproblem, and its eigenvalue need not be zero.
\begin{equation}
H\begin{pmatrix} c \\ 0 \end{pmatrix} = \begin{pmatrix} H_c\,c \\ V c \end{pmatrix}.
\end{equation}

A simple counterexample to unrestricted protection is a common detuning $\delta$ on all central orbitals. Every ideal dark vector remains compact but moves from zero to $\varepsilon = \delta$. Thus compactness alone does not pin an eigenvalue. For the $N = 2$ case, real central mixing $h$ similarly moves the antisymmetric dark state to $-h$ when $\delta = 0$.

For any finite bipartite model, $n_0 \geq |n_A - n_B|$ survives hopping changes that preserve the block-off-diagonal structure. Such changes may alter the rank-one interface condition, the localized basis and its OAM content. A proof of the lower bound therefore does not prove invariance of a selected OAM state.

\subsection{Hamiltonian of the physical waveguide array}
\label{sec:S24}

For each approximately independent polarization, the main-text geometry contains 20 Gaussian-guide modes and an OAM doublet at the hollow interface. Projecting onto these modes gives a 22-dimensional model when other radial and radiation modes can be neglected. A two-polarization description gives 44 basis states under the same truncation. The four near-gap modes described in main-text Fig.~3(b) are then two relevant interface modes per polarization; this does not require all four to be exactly degenerate.

The physical model is $H_{\mathrm{physical}} = H_{\mathrm{ideal}} + D + H_{\mathrm{mix}} + H_{\mathrm{extra}}$. $D$ contains residual onsite detunings, $H_{\mathrm{mix}}$ acts within the central doublet, and $H_{\mathrm{extra}}$ contains couplings omitted from the nearest-neighbor model, including any polarization or higher-mode effects retained in the basis. Section~\ref{sec:S4} discusses how these terms split the interface doublet.

The gap labels refer to edge-to-edge waveguide separations: $\mathrm{gap}_1$ is the hollow-to-nearest-Gaussian spacing, $\mathrm{gap}_2$ is one bulk bond spacing and $\mathrm{gap}_3$ is the alternating bulk bond spacing. For a circular Gaussian guide of radius $r_{\mathrm{G}}$ and hollow guide of outer radius $r_{\mathrm{out}}$, the corresponding center separation is $r_{\mathrm{out}} + r_{\mathrm{G}} + \mathrm{gap}_1$. For equal Gaussian guides it is $2r_{\mathrm{G}} + \mathrm{gap}_2$ or $2r_{\mathrm{G}} + \mathrm{gap}_3$. Coupling strengths depend on modal profiles as well as separations.

A compact physical-array representation separates the central block $H_c$ from the two arm matrices $H_{\mathrm{L}}$ and $H_{\mathrm{R}}$. In the ordering (central orbitals, left arm, right arm), the full Hamiltonian is
\begin{equation}
H_{\mathrm{physical}} = \begin{pmatrix}
H_c & V_{\mathrm{L}}\Tr & V_{\mathrm{R}}\Tr \\
V_{\mathrm{L}} & H_{\mathrm{L}} & 0 \\
V_{\mathrm{R}} & 0 & H_{\mathrm{R}}
\end{pmatrix}.
\end{equation}
For the ideal model, $H_c = 0$ and each $V_s$ has only its first row nonzero, equal to $(\kappa_s,\ldots,\kappa_s)$. Each arm has alternating off-diagonal entries $t_2, t_1, \ldots, t_2$. Onsite shifts, additional couplings and polarization mixing can then be inserted explicitly rather than changing the definition of the basis.

For an isolated pair of Gaussian guides with onsite difference $\Delta\delta$ and off-diagonal coupling $g$, the supermode splitting is $\sqrt{(\Delta\delta)^2 + 4|g|^2}$. At phase matching, $|g|$ is half that splitting. For a phase-matched Gaussian guide coupled equally to two unmixed OAM orbitals, the bright Gaussian--OAM splitting is $2\sqrt{2}|\kappa|$, accompanied by one dark state.
\begin{equation}
|g| = \tfrac{1}{2}\sqrt{(\beta_{\mathrm{high}} - \beta_{\mathrm{low}})^2 - (\Delta\delta)^2}.
\end{equation}

\section{Finite chains and numerical mode analysis}
\label{sec:S3}

\subsection{Chain length and interface localization}
\label{sec:S31}

The physical geometries and the eigenstate profile in main-text Fig.~2(d) use five dimers on each side. The propagation comparison in Figs.~2(f) and 2(g) uses longer arms, as specified in Sec.~\ref{sec:S53}. In the ideal bright-sector solution of Sec.~\ref{sec:S22}, the amplitude falls by $|t_1/t_2|$ per dimer. Its amplitude localization length, measured in dimers, is $\xi = 1/\ln|t_2/t_1|$ when $|t_1| < |t_2|$. The finite sum in the central-weight expression becomes a convergent geometric series as $M$ increases. This provides an analytical way to assess the contribution of the missing tail beyond a finite array.

In the finite chiral model of Sec.~\ref{sec:S22}, the zero modes are exact at any finite $M$, so finite size changes the central weight but does not split the doublet.

Define $q = |t_1/t_2|$ and $A = N\sum_s|\kappa_s/t_2|^2$. Summing the geometric series gives a closed form for the finite-chain central weight. The $q = 1$ expression is the continuous limit of the same sum.
\begin{align}
W(M) &= \left[1 + A\,\frac{1 - q^{2M}}{1 - q^2}\right]^{-1}, \qquad q \neq 1, \label{eq:S23}\\
W(M) &= (1 + AM)^{-1}, \qquad q = 1.
\end{align}
For $q < 1$, the infinite-chain central weight is $W_\infty = [1 + A/(1 - q^2)]^{-1}$. The probability in the omitted tail $j > M$ of the normalized infinite bright state is $T_{\mathrm{tail}} = W_\infty A q^{2M}/(1 - q^2)$. The truncated normalized vector is obtained by renormalizing the retained part, giving the exact relation below.
\begin{align}
W(M) &= W_\infty/(1 - T_{\mathrm{tail}}), \\
\left[W(M) - W_\infty\right]/W_\infty &= T_{\mathrm{tail}}/(1 - T_{\mathrm{tail}}).
\end{align}
At $q = 1$ the central weight tends to zero as $1/M$; at $q > 1$ it falls exponentially with $M$. The bright solution is then weighted toward the remote ends rather than localized at the central junction. The compact dark combinations remain central in the equal-profile ideal model.

\begin{figure}[t]
\centering
\includegraphics[width=0.9\textwidth]{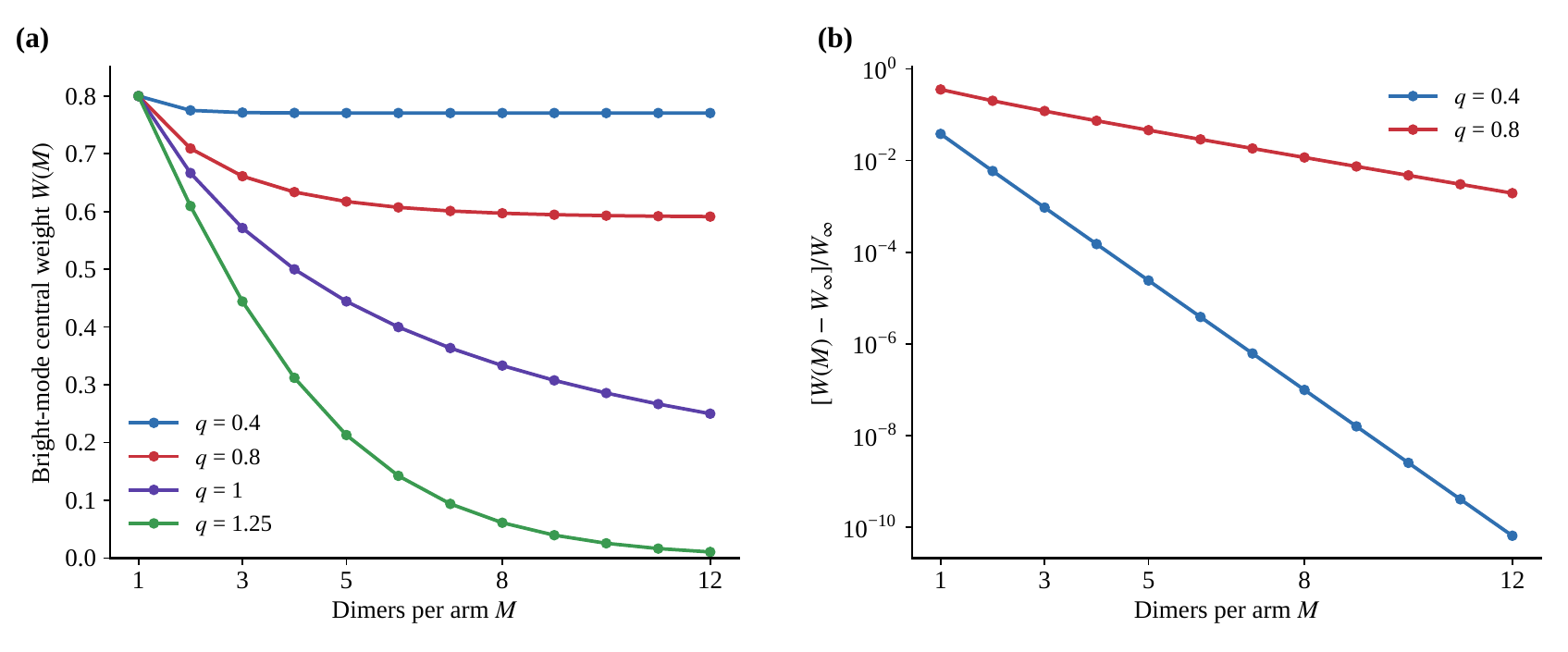}
\caption{Finite-chain localization in the explicitly defined normalized model. (a) Bright-mode central weight versus dimers per arm for four coupling ratios $q$. (b) Relative correction to the infinite-chain central weight for $q = 0.4$ and 0.8. $N = 2$, $t_2 = 1$ and $\kappa_{\mathrm{L}} = \kappa_{\mathrm{R}} = 0.25$. Curves are analytical expressions verified by independent matrix diagonalization.}
\label{fig:S1}
\end{figure}

Figure~\ref{fig:S1} evaluates these expressions for $N = 2$ and $\kappa_{\mathrm{L}}/t_2 = \kappa_{\mathrm{R}}/t_2 = 0.25$. Direct diagonalization for $M = 1,\ldots,12$ and $q = 0.4$, 0.8, 1 and 1.25 agrees with the analytical weights to $4.5\times10^{-16}$. The bright mode is selected by projecting the central symmetric orbital into the zero eigenspace; an arbitrary numerical basis within the degenerate eigenspace would mix bright and dark states.

At $q = 0.4$, $W(5) = 0.7706607362$ and $W_\infty = 0.7706422018$, with relative difference $2.4051\times10^{-5}$; the amplitude localization length is 1.0914 dimers. This verifies that five dimers suffice for that normalized example.

The central weight $W$ in this section refers to the normalized bright eigenstate. It differs from the pure-central-OAM eigenstate weight $W_c$ plotted in main-text Figs.~2(d) and 2(e); their relation is derived in Sec.~\ref{sec:S52}.

\subsection{Geometry and eigenmode calculations}
\label{sec:S32}

The physical modes are computed from the full vector Maxwell eigenproblem for modes with $\exp(i\beta z)$ dependence; the effective index is $\neff = \beta/k_0$. The interface modes are quasi-linearly polarized (see below), which justifies the fixed-polarization scalar description of Sec.~\ref{sec:S21} for each polarization.

For the S-interface of main-text Fig.~3, modes are calculated with the full-vectorial finite-difference eigensolver of Lumerical MODE at $\lambda = 1.55$\um, and the propagation in main-text Fig.~4 is simulated with Lumerical FDTD. The eigensolver returns four interface modes within the bulk gap. Their TE polarization fractions are 0 or 100\% and their waveguide TE/TM fractions exceed 99.9\%, so they form two quasi-linearly polarized doublets. The $x$-polarized doublet has $\neff = 1.440114$ and 1.440107, with effective areas of 1056 and 543\,$\mu$m$^2$, and the $y$-polarized doublet has $\neff = 1.440122$ and 1.440116, with effective areas of 670 and 1086\,$\mu$m$^2$. In each doublet, one mode is confined to the hollow waveguide and the other extends into the arms, as expected for the dark and bright zero modes of Sec.~\ref{sec:S22}. These values give splittings $\Delta\neff \approx 7\times10^{-6}$ ($x$) and $6\times10^{-6}$ ($y$), or characteristic OAM conversion lengths $\LB = \lambda/(2\Delta\neff) \approx 110$ and 130~mm. The splitting is sensitive to the computational domain and mesh; the refined calculations of Sec.~\ref{sec:S7} give $1.1\times10^{-5}$ ($x$) and $1.2\times10^{-5}$ ($y$) on the finest mesh, or $\LB \approx 70$ and 65~mm.

Geometric parameters: the hollow waveguide has inner and outer diameters $D_1 = 12$\um\ and $D_2 = 20$\um, and the Gaussian waveguides have a radius of 3.55\um. Measured edge to edge, $\mathrm{gap}_1 = \mathrm{gap}_3 = 16$\um\ and $\mathrm{gap}_2 = 8$\um, with five pairs of Gaussian waveguides on each side. The core index is 1.442 and the background index is 1.440.

FDTD settings: the simulation region is 440\um\ $\times$ 50\um\ in the transverse plane, with perfectly matched layers on all boundaries, and extends over the propagation length along $z$. The mesh is uniform, with steps of 1.5\um\ along $x$ and $y$ and 0.18\um\ along $z$, and uses conformal mesh refinement (conformal variant 0); the time step is 0.586~fs (stability factor 0.99). Each OAM input is formed by two mode sources that inject, along $+z$, two $y$-polarized first-order modes of the hollow waveguide with equal amplitudes and a relative phase of $\pm90^\circ$.

Section~\ref{sec:S7} checks the convergence of the doublet splitting with an independent mode solver and a beam-propagation calculation.

\subsection{Propagation-constant matching}
\label{sec:S33}

Propagation-constant matching refers to the local modes of the constituent waveguides at the same wavelength and material parameters. The ideal reference takes $\beta_{\mathrm{G}} = \beta_{+l} = \beta_{-l} = \beta_0$. This combines degeneracy of the opposite-charge OAM pair with matching between that pair and the Gaussian mode. Because $\beta = k_0\neff$, matching can be sought by adjusting waveguide dimensions to bring the effective indices together. It does not require identical transverse profiles or equal propagation constants for every eigenmode of the coupled array.

Writing $\delta_{\mathrm{G},l} = \beta_{\mathrm{G}} - \beta_l$ for the isolated-guide mismatch, we first match the isolated Gaussian and hollow-guide modes by adjusting the Gaussian-waveguide radius and then examine the full-array eigenmodes. Matching the isolated propagation constants does not by itself make the array doublet degenerate, because the Gaussian waveguides perturb the two standing-wave orbitals of the hollow waveguide differently (Sec.~\ref{sec:S4}).

For a geometric parameter $p$ near a trial value $p_0$, a first-order correction to the isolated-guide mismatch is $\delta_{\mathrm{G},l}(p) \simeq \delta_{\mathrm{G},l}(p_0) + [\partial_p\beta_{\mathrm{G}} - \partial_p\beta_l]_{p_0}(p - p_0)$. If only the Gaussian radius is varied, the derivative of the isolated hollow mode is zero. A nonzero slope gives the local crossing estimate below, which can be checked with one additional eigenmode calculation.
\begin{equation}
p^* \simeq p_0 - \frac{\delta_{\mathrm{G},l}(p_0)}{\left[\partial_p\delta_{\mathrm{G},l}\right]_{p_0}}.
\end{equation}

\subsection{Higher-order OAM modes}
\label{sec:S34}

Main-text Fig.~3(e) shows a separate larger hollow-waveguide geometry associated with $l = \pm3$. The relevant even and odd angular functions are proportional to $\cos(3\varphi)$ and $\sin(3\varphi)$; compatible quadrature superpositions yield $\exp(\pm 3i\varphi)$. Its radial profiles and propagation constants are computed separately from those of the $l = \pm1$ geometry.

This example illustrates a higher-order pair in another structure, not simultaneous propagation of the $\pm1$ and $\pm3$ families in one device.

For any nonzero integer $l$, choose angular functions $f_c = R(r)\cos(l\varphi)/\sqrt{\pi}$ and $f_s = R(r)\sin(l\varphi)/\sqrt{\pi}$ with $\int|R|^2 r\,dr = 1$ over the chosen annular interval. They are orthonormal on that region, and $(f_c \pm if_s)/\sqrt{2} = R(r)\exp(\pm il\varphi)/\sqrt{2\pi}$. This identity proves the OAM construction for $l = 3$ as well as $l = 1$, provided the two radial functions are the same.

When the radial profiles differ, the angular Fourier decomposition in Sec.~\ref{sec:S35} provides the appropriate modal fractions.

\subsection{Propagation of the OAM superpositions}
\label{sec:S35}

In the ideal balanced doublet, orthogonal standing-wave modes $u_e$ and $u_o$ can be combined as $u_\pm = (u_e \pm iu_o)/\sqrt{2}$. This construction yields local OAM only when the annular parts have compatible radial dependence and appropriate relative weights. If their local weights differ, the excitation coefficients must be adjusted to produce the desired annular field.
\begin{equation}
E(z) = c_e u_e\exp(i\beta_e z) + c_o u_o\exp(i\beta_o z).
\end{equation}
For an isolated lossless doublet and the fixed balanced superposition basis, excitation of $u_+$ produces the following conversion probabilities.
\begin{equation}
P_+(z) = \cos^2(\Delta\beta\,z/2), \qquad P_-(z) = \sin^2(\Delta\beta\,z/2).
\end{equation}
Here $\Delta\beta = \beta_e - \beta_o$, so near-preservation on a length $L$ requires $|\Delta\beta|L \ll 1$ in this idealized description. A field launched only into the central guide generally also excites bulk components; a two-mode expansion alone does not reproduce that full initial condition.

For the complete lossless finite coupled-mode model, use spectral propagation $a(z) = U\exp(i\Lambda z)U\Tr a(0)$, where $H = U\Lambda U\Tr$. For the physical array, propagation is computed directly with FDTD (Sec.~\ref{sec:S32}), and main-text Fig.~4 shows cross sections at 0, 4 and 15~mm.

For a scalar fixed-polarization field in a specified annular region $\mathcal{A}$, define the local power fraction $\eta_{\mathcal{A}} = \int_{\mathcal{A}}|E|^2dA / \int_{\mathrm{all}}|E|^2dA$. At each radius, the angular coefficient is $c_m(r,z) = (2\pi)^{-1}\int_0^{2\pi}E(r,\varphi,z)\exp(-im\varphi)\,d\varphi$. Integrating $|c_m|^2 r$ over the selected radial interval and normalizing over angular orders gives an OAM fraction $p_m$ within that region. A high conditional OAM fraction $p_m$ does not by itself imply a high retained power $\eta_{\mathcal{A}}$. Vector modes require an explicitly stated power convention.

The unequal-weight case can also be completed analytically. Suppose the annular restrictions of normalized global eigenmodes are $u_e|_{\mathcal{A}} = a f_c$ and $u_o|_{\mathcal{A}} = b f_s$, where $a, b > 0$ after choosing consistent phases, and $f_c, f_s$ are the normalized local angular functions above. For $c_e = x$ and $c_o = iy$ with $x, y$ real, write $X = ax$ and $Y = by$. The local field is $Xf_c\exp(i\beta_e z) + iYf_s\exp(i\beta_o z)$. Its projection amplitudes onto the local $\pm l$ basis are
\begin{gather}
A_\pm(z) = \left[X\exp(i\beta_e z) \pm Y\exp(i\beta_o z)\right]/\sqrt{2}, \\
p_\pm(z) = \frac{1}{2} \pm \frac{XY\cos(\Delta\beta z)}{X^2 + Y^2}.
\end{gather}
Thus equal global excitation coefficients do not in general create a pure local OAM field. Local purity at $z = 0$ requires $X = Y$, giving $c_o/c_e = ia/b$. With globally normalized input and orthogonal global eigenmodes, $|c_e|^2 = b^2/(a^2 + b^2)$ and $|c_o|^2 = a^2/(a^2 + b^2)$. The annular fraction of this matched two-mode state is then
\begin{equation}
\eta_{\mathcal{A}} = \frac{2a^2b^2}{a^2 + b^2}.
\end{equation}
These equations separate the power retained near the interface from its conditional OAM composition. If the radial profiles are not matched, $a$ and $b$ alone are insufficient and the full angular decomposition is required. A phase winding of $l$ around a contour also does not, by itself, determine the fraction of power in angular order $l$.

For the balanced case, requiring $p_-(z) \leq \epsilon$ throughout $0 \leq z \leq L$ gives the sufficient first-lobe bound $|\Delta\beta|L \leq 2\arcsin\sqrt{\epsilon}$. For $\epsilon = 0.01$ and $L = 15$~mm, it is $|\Delta\beta| \leq 0.0133557$~mm$^{-1}$. This is an analytical design criterion. An endpoint-only criterion would allow revivals and would not ensure preservation at all intermediate distances. For the MODE splitting of the $x$-polarized doublet (Sec.~\ref{sec:S32}), $\Delta\beta = k_0\Delta\neff \approx 0.028$~mm$^{-1}$ and $\LB = \pi/\Delta\beta \approx 110$~mm, and the balanced two-mode estimate gives $p_-(15~\mathrm{mm}) = \sin^2(\Delta\beta L/2) \approx 0.045$. For the finest-grid splitting of the $y$-polarized doublet excited in main-text Fig.~4 (Sec.~\ref{sec:S7}), $\Delta\beta \approx 0.049$~mm$^{-1}$ and $\LB \approx 65$~mm, which gives $p_-(15~\mathrm{mm}) \approx 0.13$, close to the value 0.14 from the beam-propagation calculation of Sec.~\ref{sec:S7}. Both exceed the 1\% criterion above; because $p_+$ remains larger than $p_-$, the phase winding keeps the sign of the input, as in main-text Fig.~4.

A split doublet does not inevitably mix OAM: if its eigenvectors themselves have definite opposite OAM, diagonal detuning only changes their phases. The conversion formula used here applies to the standing-wave pair that is combined into local OAM, as described in the main text. More generally, for a constant effective Hamiltonian in an orthonormal OAM basis, the off-diagonal coupling determines conversion.
\begin{gather}
H_2 = \bar\beta I + \begin{pmatrix} d & g \\ g^* & -d \end{pmatrix}, \\
P_{+\to-}(z) = \frac{|g|^2}{|g|^2 + d^2}\sin^2\!\left[z\sqrt{|g|^2 + d^2}\right].
\end{gather}
This follows from $(H_2 - \bar\beta I)^2 = (|g|^2 + d^2)I$ when $d$ is real. A zero off-diagonal $g$ gives no conversion even when $d$ is nonzero.

For a general initial field, the coefficients in a complete orthonormal mode basis are $c_n = \langle u_n, E(0)\rangle$. Summing $c_n u_n\exp(i\beta_n z)$ conserves the norm in a lossless model. The bulk components do not disappear globally: they can spread away from the central observation region and can recur in a finite system. This is why initial spreading in main-text Fig.~4 is not evidence of dissipative relaxation into a pure eigenstate.

The angular decomposition also gives the mean scalar angular order within the annulus as $\sum_m m\,p_m$. If only $\pm l$ are present, this is $l(p_{+l} - p_{-l})$. In the balanced standing-wave model it oscillates as $l\cos(\Delta\beta z)$.

\subsection{OAM retention after a central-only launch}
\label{sec:S36}

Consider the lossless $N = 2$ equal-profile Hamiltonian of Sec.~\ref{sec:S22} with $h = 0$ and no onsite detuning. Work in an orthonormal single-polarization basis. Write $b = (c_+ + c_-)/\sqrt{2}$ and $d = (c_+ - c_-)/\sqrt{2}$. The compact $d$ is a zero mode, and the normalized extended bright zero mode has central weight $W$. A central-only launch $c_+ = (b + d)/\sqrt{2}$ has total zero-subspace overlap $(1 + W)/2$ and is not itself an eigenstate.

Let $f(z) = \langle b|\exp(iHz)|b\rangle$, consistent with the propagation sign in Sec.~\ref{sec:S21}. In the bright sector, nonzero eigenvalues occur in pairs $\pm\varepsilon$, with equal weight $w_\varepsilon$ on $b$. The spectral resolution therefore gives
\begingroup
\addtocounter{equation}{-1}
\renewcommand{\theHequation}{sub.\arabic{equation}}
\begin{subequations}
\begin{equation}
f(z) = W + 2\sum_{\varepsilon>0}w_\varepsilon\cos(\varepsilon z), \qquad 2\sum_{\varepsilon>0}w_\varepsilon = 1 - W.
\end{equation}
All $w_\varepsilon$ are nonnegative, so $f$ is real and lies between $2W - 1$ and 1 at every propagation distance. Taking the overlap with $c_+$ and $c_-$ gives
\begin{gather}
c_+(z) = [1 + f(z)]/2, \qquad c_-(z) = [f(z) - 1]/2, \\
P_c(z) = [1 + f(z)^2]/2, \qquad F_+(z) = \frac{1}{2} + \frac{f(z)}{1 + f(z)^2}. \label{eq:S34c}
\end{gather}
\end{subequations}
\endgroup
Here $P_c$ denotes the total central modal power for unit input power, and $F_+$ denotes the conditional positive-charge fraction, written $\Fp$ in main-text Eq.~(8). The bright-mode weight $W$ differs from $W_c$. Since $1/2 + f/(1 + f^2)$ is nondecreasing on $-1 \leq f \leq 1$, the lower bound $f \geq m = 2W - 1$ gives $F_+ \geq 1/2 + m/(1 + m^2)$. If $W \geq 1/2$, $f$ is nonnegative and $P_c \geq (1 + m^2)/2$. If $W < 1/2$, the interval only gives $P_c \geq 1/2$.

For the normalized example $N = 2$, $M = 5$, $t_1/t_2 = 0.4$ and $\kappa_{\mathrm{L}}/t_2 = \kappa_{\mathrm{R}}/t_2 = 0.25$, $W = 0.7706607362$, so $F_+ \geq 0.9186460620$ and $P_c \geq 0.6465144682$ at every distance in this ideal model.

The stationary preparation is different. Let $\Psi_B$ have central component $\sqrt{W}\,b$ and let $D = d$. The normalized state $(\Psi_B + \sqrt{W}D)/\sqrt{1 + W}$ is an exact zero mode with a pure positive-charge central field and central weight $W_c = 2W/(1 + W)$, as derived explicitly in Sec.~\ref{sec:S52}. Its required arm tails are absent from a central-only input. Thus this construction and the dynamical bound address different excitation protocols.

Bond perturbations that preserve chiral symmetry and the equal-profile rank-one interface leave the reasoning applicable with the corresponding $W$. Perturbations that introduce central mixing or change the coupling rank invalidate the return-amplitude formula (Sec.~\ref{sec:S6}).

\section{W-interface and comparison with the S-interface}
\label{sec:S4}

The main text labels the geometry with $\mathrm{gap}_1$ and $\mathrm{gap}_3$ greater than $\mathrm{gap}_2$ as the S-interface. It labels the configuration with $\mathrm{gap}_1$ and $\mathrm{gap}_3$ smaller than $\mathrm{gap}_2$ as the W-interface and associates it with non-negligible $h$ in main-text Eq.~(3).

Comparing the interface spectrum, mode profiles, central weight and doublet splitting of the two geometries separates changes in localization from changes in OAM coherence. Direct coupling between the central orbitals breaks the chiral structure used in Sec.~\ref{sec:S22}.

In a local doublet basis with onsite detunings $\delta_+$ and $\delta_-$ and mixing $h$, the local splitting is $2\sqrt{|h|^2 + [(\delta_+ - \delta_-)/2]^2}$. This is a local two-state result; the actual array splitting may also contain shifts from coupling to surrounding modes.

Larger splitting can cause an initially prepared OAM superposition to evolve appreciably within the observation length even if both contributing eigenmodes remain localized. Conversely, localization can change while a compact interference state remains. Neither the zero-mode count nor a central intensity image therefore determines whether the OAM field is maintained.

The local doublet splitting quoted above follows by solving $\det(H_c - \varepsilon I) = 0$, with $\bar\delta = (\delta_+ + \delta_-)/2$ and $d = (\delta_+ - \delta_-)/2$. The eigenvalues are $\bar\delta \pm \sqrt{d^2 + |h|^2}$. The surrounding array can change this splitting. To see the dependence explicitly, eliminate the arm amplitudes $y$ from $Vc + H_{\mathrm{arms}}y = \varepsilon y$, whenever $\varepsilon I - H_{\mathrm{arms}}$ is invertible.
\begin{align}
y &= (\varepsilon I - H_{\mathrm{arms}})^{-1}Vc, \\
H_{\mathrm{eff}}(\varepsilon) &= H_c + V\Tr(\varepsilon I - H_{\mathrm{arms}})^{-1}V.
\end{align}
Because of this energy-dependent correction, the splitting of an isolated annulus differs from the full-array doublet splitting. Near an arm resonance, the inverse is singular and the reduced formula must be replaced by the full eigenproblem.

For the ideal $N = 2$ equal-profile system, let $|\psi_B\rangle$ be the normalized bright zero mode with central weight $W$ and let $|D\rangle$ be the compact antisymmetric dark state. Add a common central detuning $\delta$ and a real central mixing $h$. In the basis $(|\psi_B\rangle, |D\rangle)$, the first-order projected perturbation is
\begin{gather}
P\,\delta H\,P = \begin{pmatrix} W(\delta + h) & 0 \\ 0 & \delta - h \end{pmatrix}, \\
\varepsilon_B - \varepsilon_D \simeq (W - 1)\delta + (W + 1)h. \label{eq:S38}
\end{gather}
The dark eigenvalue $\varepsilon_D = \delta - h$ is exact because the dark vector decouples from the arms. The bright shift is first order. For this perturbation restricted to the central A sublattice, the second-order correction vanishes: the reduced inverse of the ideal chiral Hamiltonian maps A to B, while the central perturbation annihilates B. Hence the leading omitted eigenvalue correction is of order $\|H_c\|^3/\Delta^2$ for $\|H_c\|$ much smaller than the finite spectral separation $\Delta$. General perturbations outside this central block need not have this cancellation. Other central matrices can also mix the projected states.

\begin{figure}[t]
\centering
\includegraphics[width=0.9\textwidth]{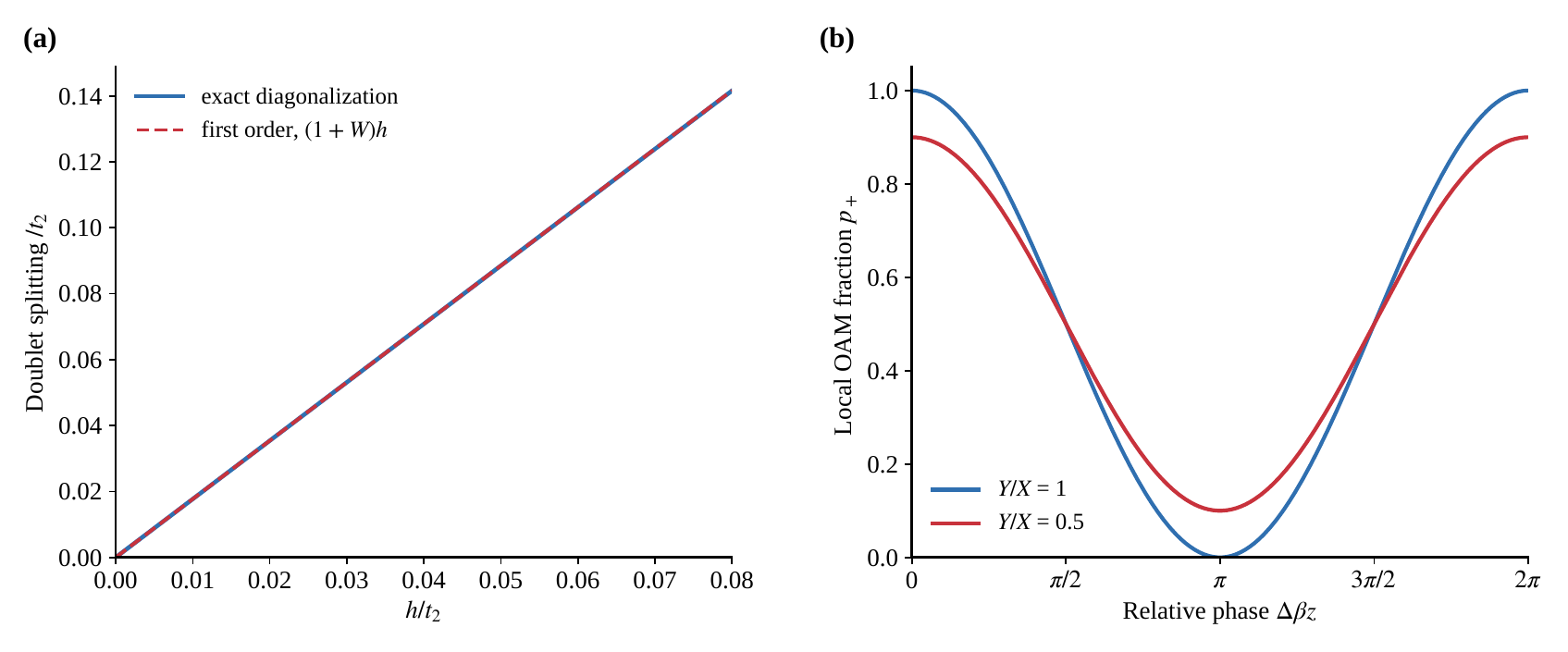}
\caption{Doublet splitting and OAM composition in explicit model calculations. (a) Exact finite-matrix splitting and the first-order prediction $(1 + W)h$ for $N = 2$, $M = 5$, $t_1/t_2 = 0.4$, $\kappa_{\mathrm{L}}/t_2 = \kappa_{\mathrm{R}}/t_2 = 0.25$ and $\delta = 0$. (b) Local conditional $p_+$ versus relative phase for balanced and unequal angular weights.}
\label{fig:S2}
\end{figure}

For the normalized baseline of Sec.~\ref{sec:S31}, $W = 0.7706607362$. At $h/t_2 = 10^{-5}$ and $\delta/t_2 = 2\times10^{-5}$, the exact signed splitting is $1.3119822081\times10^{-5}\,t_2$, agreeing with the first-order expression to $3.9\times10^{-15}\,t_2$. Figure~\ref{fig:S2}(a) shows the mixing-only comparison over a wider interval.

In the physical array, direct mixing arises even when the modes of the isolated hollow waveguide are degenerate. The Gaussian waveguides lie along the array axis, so they perturb the standing-wave orbital whose lobes point along this axis more strongly than the orthogonal orbital and shift it by an amount $\varepsilon$ relative to it. In the $\pm l$ basis, this differential shift is a common detuning $\varepsilon/2$ together with a mixing $h = \varepsilon/2$, and by Eq.~\eqref{eq:S38} it splits the doublet by $W\varepsilon$. Chiral symmetry does not suppress this term. A mismatch $\delta_c$ between the hollow and Gaussian waveguides adds $(W - 1)\delta_c$ to the same splitting, so the two contributions cancel for $\delta_c = W\varepsilon/(1 - W)$. The residual splitting of the physical doublet can therefore be reduced by adjusting the Gaussian-waveguide radius, at the cost of a small deliberate mismatch; Sec.~\ref{sec:S7} demonstrates this for the physical array.

In the outward-arm description, the stronger first Gaussian--Gaussian bond of the S termination corresponds to $|t_1/t_2| < 1$ when overlap decreases with distance; reversing that bond pattern in W can give $|t_1/t_2| > 1$. The stronger central contact of W also changes $\kappa_s$, but cannot by itself reverse the arm decay ratio or prove a nonzero $h$. The two mechanisms are therefore distinct.

For illustration at equal central coupling $\kappa_{\mathrm{L}} = \kappa_{\mathrm{R}} = 0.25J$ and $M = 5$, choose S with $(t_1, t_2) = (0.4J, J)$ and W with $(t_1, t_2) = (J, 0.4J)$. The bulk dispersions coincide, with inner band edges $\pm0.6J$ and outer band edges $\pm1.4J$, but Eq.~\eqref{eq:S16} gives central bright weights 0.7706607362 and 0.0003522344, respectively. The exact finite spectra have nearest nonzero magnitudes $0.6703856884J$ and $0.0086052479J$. Therefore the smaller finite-array separation in the W example, rather than the bulk band edge alone, controls the validity of isolated-doublet perturbation theory.

These normalized parameters isolate the effect of reversing the arm termination at fixed contact coupling. The physical S and W geometries also differ in contact separation, so this comparison isolates the mechanism rather than reproducing the full-wave comparison. Even the W ideal model permits a pure local OAM stationary superposition, but its central power fraction $2W/(1 + W)$ is only 0.0007042207 in this example.

\section{Comparison of isolated and coupled OAM waveguides}
\label{sec:S5}

\subsection{Common model and numerical parameters}
\label{sec:S51}

Main-text Fig.~2 compares one isolated hollow guide with the same central OAM doublet coupled to either uniform or SSH Gaussian-waveguide arms, together with two further references built on the uniform arms (Sec.~\ref{sec:S54}). The coupled models have $N = 2$ central orbitals and dimension $2 + 4M$; each arm contains $M$ dimers with the outward ordering defined in Sec.~\ref{sec:S22}. All calculations use an orthonormal scalar basis for one polarization, matched onsite propagation constants (except for the detuned reference), zero direct central OAM mixing, and equal real contacts $\kappa_{\mathrm{L}} = \kappa_{\mathrm{R}} = \kappa$. Power means the conserved modal norm in this model.

Let $J$ denote an inverse-length coupling scale. The SSH parameters are $t_2/J = 1$, $t_1/J = 0.4$, and $\kappa/J = 0.25$; the uniform reference has $t_2/J = t_1/J = 0.7$ with the same $\kappa/J$. Thus both coupled references have the same contact strength and $t_1 + t_2 = 1.4J$, although their bulk gaps differ. The isolated reference has no arms and $\kappa = 0$. The strongly coupled reference uses the uniform arms with $\kappa/J = 1.2$, and the detuned reference uses the uniform arms with $\kappa/J = 0.25$ and a common onsite detuning $\delta/J = 3$ of both central orbitals. These dimensionless parameters specify the comparison and are not fitted electromagnetic parameters for main-text Figs.~3 and 4.

The propagation generator $H$ used here is measured relative to the common propagation constant. Its relation to the physical generator and an array eigenvalue is
\begin{equation}
H_{\mathrm{phys}} = \beta_0 I + H, \qquad H|\Psi\rangle = (\beta - \beta_0)|\Psi\rangle.
\end{equation}
A zero eigenvalue of $H$ therefore corresponds to $\beta = \beta_0$, not to a vanishing physical propagation constant. Matching removes the common diagonal term; zero eigenvalues additionally require a state that satisfies the coupled-site equations. Nonzero eigenvalues of $H$ remain present in the same array.

\subsection{Pure-central-OAM eigenstates and their normalization}
\label{sec:S52}

Choose $c_+ = C$ and $c_- = 0$. The opposite-charge construction is obtained by interchanging these amplitudes. The complete eigenstate also includes Gaussian-site amplitudes on both arms:
\begin{equation}
b_{s,j} = 0, \qquad a_{s,j} = -\frac{\kappa_s}{t_2}\left(-\frac{t_1}{t_2}\right)^{j-1}C.
\label{eq:S40}
\end{equation}
Substitution verifies every component of $H|\Psi_+\rangle = 0$. The central and A-site equations vanish because all B-site amplitudes are zero. At the first B site, $\kappa_s C + t_2 a_{s,1} = 0$; at subsequent B sites, $t_1 a_{s,j-1} + t_2 a_{s,j} = 0$. The outer A-site equation is also satisfied by the specified finite termination. Thus the zero eigenvalue follows from cancellation of the coupled amplitudes. When $|t_1/t_2| < 1$, these Gaussian-site amplitudes decay geometrically away from the interface. They are the transverse tails of the stationary eigenstate, not a longitudinal wake or a continuing loss of power.

Define $q = |t_1/t_2|$. Summing the normalized central power and all arm powers yields the pure-central-charge eigenstate weight:
\begin{equation}
W_c(M) = |C|^2 = [1 + A(M)]^{-1}, \qquad A(M) = \sum_{s=\mathrm{L},\mathrm{R}}\left|\frac{\kappa_s}{t_2}\right|^2\sum_{j=0}^{M-1}q^{2j}.
\label{eq:S41}
\end{equation}
For equal real contacts, $A(M) = 2(\kappa/t_2)^2(1 - q^{2M})/(1 - q^2)$ when $q \neq 1$, and $A(M) = 2M(\kappa/t)^2$ for uniform arms with $t_1 = t_2 = t$. The resulting limiting expressions reproduce main-text Eqs.~(6) and (7):
\begin{align}
W_c^{\mathrm{SSH}}(\infty) &= \left[1 + \frac{2\kappa^2}{t_2^2 - t_1^2}\right]^{-1}, \\
W_c^{\mathrm{uniform}}(M) &= \left[1 + \frac{2M\kappa^2}{t^2}\right]^{-1}. \label{eq:S43}
\end{align}
For the parameters above, $W_c(5) = 0.8704781446$ for SSH and 0.4394618834 for the uniform reference, while the isolated value is one. The SSH limit is 0.8704663212. The uniform pure-charge construction is normalizable for every finite $M$, but its central weight vanishes as $M$ grows; it does not approach a localized, normalizable pure-central-charge zero mode on infinite uniform arms.

The distinction from the bright-state weight $W$ in Eqs.~\eqref{eq:S16} and \eqref{eq:S23} is essential. For $N = 2$, a unit bright central amplitude drives each arm with $\sqrt{2}\kappa_s$, giving $W = [1 + 2A(M)]^{-1}$. Consequently,
\begin{equation}
W_c = \frac{2W}{1 + W}, \qquad W = \frac{W_c}{2 - W_c}.
\end{equation}
For example, the five-dimer SSH values $W = 0.7706607362$ and $W_c = 0.8704781446$ describe different normalized eigenstates of the same matrix, not inconsistent estimates of one quantity. The compact dark state $(|{+l}\rangle - |{-l}\rangle)/\sqrt{2}$ has unit central weight in both coupled references but contains both OAM charges. Combining that state with the extended bright zero mode gives the pure-central-charge state described in Sec.~\ref{sec:S36}. The compact state alone therefore does not establish preservation of a locally launched single charge.

In main-text Fig.~2(d), $p_n$ is power per physical waveguide: $p_n = |a_n|^2$ on a scalar Gaussian site and $p_0 = |c_+|^2 + |c_-|^2$ on the hollow guide. The spatial profiles use $M = 5$ and sum to one over all physical sites. Panel (e) evaluates Eqs.~\eqref{eq:S41}--\eqref{eq:S43} for $M = 1,\ldots,40$. The plotted central weights are $W_c$, not $W$.

\subsection{Local excitation and propagation observables}
\label{sec:S53}

For propagation, each system receives unit amplitude in its central $+l$ orbital, with all other amplitudes zero. This common local initial condition does not include the Gaussian tails of Eq.~\eqref{eq:S40}. Diagonalize the dimensionless Hermitian matrix $H/J$ as $U\Lambda U\Tr$ and evaluate the spectral propagator at $\tau = Jz$:
\begin{equation}
a(\tau) = U\exp(i\Lambda\tau)U\Tr a(0).
\end{equation}
The central power and conditional central fraction plotted in main-text Figs.~2(f) and 2(g) are
\begin{equation}
P_c = |c_+|^2 + |c_-|^2, \qquad \Fp = |c_+|^2/P_c.
\end{equation}
The conditional fraction is defined only when central power is nonzero. It is neither total retained power nor the mean OAM. In the two-orbital central basis, the conditional mean angular order is $l(2\Fp - 1)$. Gaussian-arm sites carry scalar modes and are not assigned copies of the central OAM pair.

Section~\ref{sec:S36} explains the dynamics through the bright return amplitude $f$. The absolute target power is $P_+ = |1 + f|^2/4$, while $P_c$ and $\Fp$ follow Eq.~\eqref{eq:S34c}. The local launch occupies both a compact dark component and a bright component that can excite dispersive modes. Consequently, the uniform-array central power need not fall to zero, yet its surviving central field need not retain the initial OAM composition. The SSH bright sector has substantial interface-mode weight, which limits the redistribution under the specified conditions.

The propagation calculation uses $M = 250$ per arm and 4001 equally spaced values of $\tau$ from 0 to 200. Increasing $M$ to 300 checks that end reflections do not affect the displayed curves. Spectral propagation conserves the norm without a finite-step integration approximation; the sampling interval controls the plotted resolution. The maximum norm error is below $10^{-14}$ and the $M = 250$ versus $M = 300$ difference in the reported central observables is below $10^{-12}$ for all five systems. The analytic eigenvectors at $M = 5$ have residual norm $\|H\Psi\|/J$ below $10^{-12}$.

At $\tau = 60$, the isolated reference has $P_c = P_+ = \Fp = 1$. The uniform reference gives $P_c = 0.500000$, $P_+ = 0.250238$, and $\Fp = 0.500475$; the SSH reference gives $P_c = 0.799688$, $P_+ = 0.786940$, and $\Fp = 0.984060$. These values are finite-distance outputs, not an assumption of irreversible relaxation or an exact asymptotic limit. Finite lossless arrays can exhibit recurrences at longer distances. Over $100 \leq \tau \leq 200$, the time-averaged central power is 0.797 for the SSH interface and 0.500 for the uniform reference, and $\Fp$ stays within 0.983--0.984 and at 0.500, respectively.

The comparison establishes that, for the stated model and common local input, the SSH termination maintains a finite central OAM eigenstate weight and retains more target central OAM power than the uniform reference. The isolated guide is the uncoupled benchmark, not a performance level exceeded by the SSH array. Bond disorder and unequal contacts are treated in Sec.~\ref{sec:S6}; direct OAM mixing, which this mechanism does not suppress, is discussed in Sec.~\ref{sec:S4}.

\subsection{Strongly coupled and detuned references}
\label{sec:S54}

Main-text Figs.~2(f) and 2(g) and Table~II include two further references built on the uniform arms, $t_1 = t_2 = t = 0.7J$. They test whether localization of the bright component alone retains the central charge. In both, the dark combination $(|{+l}\rangle - |{-l}\rangle)/\sqrt{2}$ remains decoupled from the arms, so the charge dynamics is set by where the localized bright component lies relative to the dark mode.

\begin{figure}[t]
\centering
\includegraphics[width=0.9\textwidth]{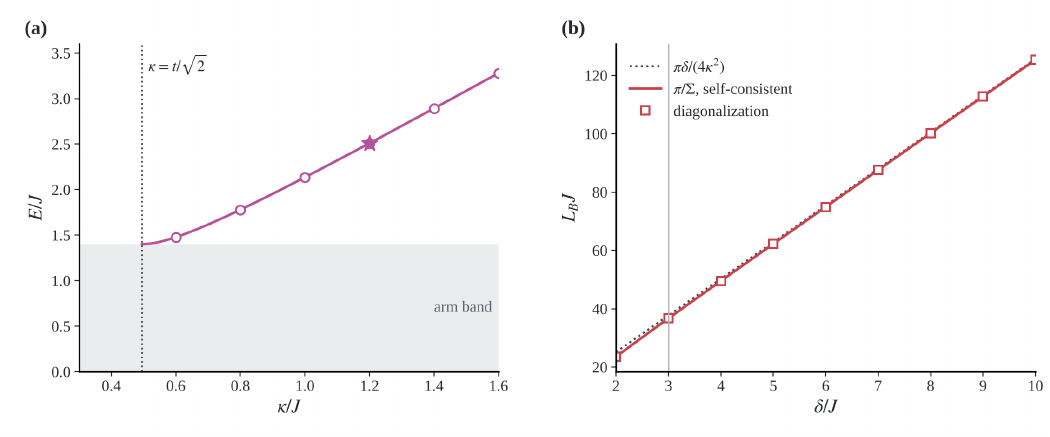}
\caption{Strongly coupled and detuned references of main-text Fig.~2. (a) Energy $E$ of the bright bound state above the arm band for uniform arms ($t = 0.7J$) versus the contact coupling $\kappa$: Eq.~\eqref{eq:S47} (line) and diagonalization with 200 dimers per arm (circles). The dotted line marks the threshold $\kappa = t/\sqrt{2}$, the shaded region the arm band, and the star the value $\kappa = 1.2J$ used in main-text Fig.~2. (b) Charge-conversion length $\LB$ of the detuned reference versus $\delta$ at $\kappa = 0.25J$: numerical bright--dark splitting (squares), self-consistent solution of Eq.~\eqref{eq:S48} (solid line) and leading order $\pi\delta/(4\kappa^2)$ (dotted line). The vertical line marks $\delta = 3J$.}
\label{fig:S3}
\end{figure}

\textit{Strong contact coupling.} The bright orbital couples to the first site of each arm with $\sqrt{2}\kappa$. Consider first a bound state above the band, $E > 2t$. Eliminating the arms, it satisfies $E = 4\kappa^2 G(E)$, where $G(E) = [E - (E^2 - 4t^2)^{1/2}]/(2t^2)$ is the end-site Green's function of a semi-infinite uniform arm on this branch. Using $t^2G^2 - EG + 1 = 0$ gives $G = (4\kappa^2 - t^2)^{-1/2}$, which describes a decaying state only when $tG < 1$, that is, for $\kappa > t/\sqrt{2}$. Chiral symmetry supplies a partner below the band at $-E$, so the bound states lie at
\begin{equation}
E = \pm\frac{4\kappa^2}{(4\kappa^2 - t^2)^{1/2}}.
\label{eq:S47}
\end{equation}
For $\kappa = 1.2J$, $E = \pm2.509J$ [Fig.~\ref{fig:S3}(a)]. The model remains chiral, and its zero modes are the dark mode and an extended bright zero mode, so the localized bright weight resides at $\pm E$ rather than at zero. The bright return amplitude $f$ of Sec.~\ref{sec:S36} therefore contains a non-decaying term proportional to $\cos(Ez)$, and $\Fp$, given by Eq.~\eqref{eq:S34c}, oscillates between nearly 0 and 1 with a half period $\pi/E = 1.25/J$.

\textit{Onsite detuning.} A common detuning $\delta$ of both central orbitals breaks chiral symmetry. The dark mode moves exactly to $\delta$ (Sec.~\ref{sec:S23}), whereas the bright bound state lies at $\delta + \Sigma$, with
\begin{equation}
\Sigma = 4\kappa^2 G(\delta + \Sigma) \approx 4\kappa^2/\delta, \qquad \delta \gg 2t.
\label{eq:S48}
\end{equation}
The two components acquire a relative phase $\Sigma z$, and the central charge converts fully from $+l$ to $-l$ over $\LB = \pi/\Sigma \approx \pi\delta/(4\kappa^2)$. For $\delta = 3J$, the numerical splitting is $\Sigma = 0.0857J$, giving $\LB = 36.7/J$; the leading-order value is $37.7/J$. Figure~\ref{fig:S3}(b) compares the numerical splitting, the self-consistent solution of Eq.~\eqref{eq:S48} and its leading order for $2J \leq \delta \leq 10J$; the numerical and self-consistent values agree to a relative accuracy better than $10^{-3}$. Because $\LB$ grows only linearly with $\delta$ at fixed $\kappa$, retaining the charge over a length $L$ by detuning requires $\delta \gg 4\kappa^2 L/\pi$, which amounts to decoupling the hollow waveguide from the array.

Over $100 \leq \tau \leq 200$, the time-averaged central power is 0.706 for the strongly coupled reference and 0.970 for the detuned reference, compared with 0.797 for the SSH interface and 0.500 for the uniform reference. The minimum conditional fractions over the same interval are 0.002, 0.000, 0.983 and 0.500, respectively. The detuned reference therefore retains more power than the SSH interface, whereas the strongly coupled reference retains less, and neither retains the charge. At $\tau = 60$, the strongly coupled reference gives $P_c = 0.886248$ and $\Fp = 0.995864$, and the detuned reference gives $P_c = 0.970159$ and $\Fp = 0.707678$; these instantaneous values sample oscillations and are not representative of retention. Both calculations use $M = 250$ per arm, converged against $M = 300$ to better than $10^{-12}$.

\begin{figure}[t]
\centering
\includegraphics[width=0.9\textwidth]{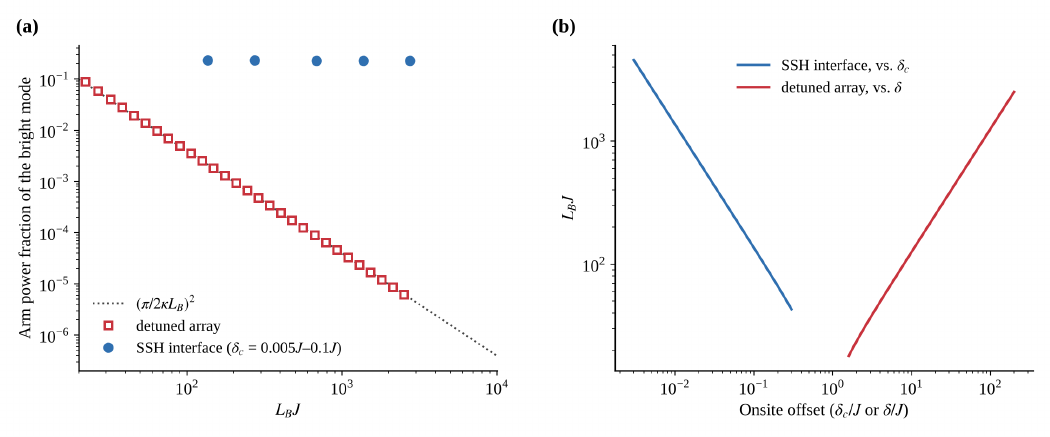}
\caption{Conversion length and hybridization of the SSH and detuned routes. (a) Fraction of the bright-mode power on the arms versus the charge-conversion length $\LB$. Squares: detuned reference (uniform arms, $t = 0.7J$, $\kappa = 0.25J$) for $1.6J \leq \delta \leq 200J$; dotted line: $(\pi/2\kappa\LB)^2$. Circles: SSH interface of Sec.~\ref{sec:S51} with a residual central mismatch $\delta_c = 0.005J$, $0.01J$, $0.02J$, $0.05J$ and $0.1J$. (b) $\LB$ versus the onsite offset: central mismatch $\delta_c$ for the SSH interface and detuning $\delta$ for the detuned reference. All calculations use $M = 300$ dimers per arm.}
\label{fig:S4}
\end{figure}

The detuned route also removes the coupling that the interface is meant to retain. Figure~\ref{fig:S4} compares the two routes at equal conversion length. At the SSH interface, a residual central mismatch $\delta_c$ splits the doublet by $(1 - W)\delta_c$ (Sec.~\ref{sec:S4}), while the bright interface mode keeps a fraction $1 - W \approx 0.23$ of its power on the arms, nearly independent of $\delta_c$. In the detuned reference, the bright bound state has an arm fraction of about $4\kappa^2/\delta^2$, which with $\LB \approx \pi\delta/(4\kappa^2)$ becomes $(\pi/2\kappa\LB)^2$. At $\LB = 685/J$, the value produced at the SSH interface by $\delta_c = 0.02J$, the detuned reference keeps $8\times10^{-5}$ of the bright-mode power on the arms, about 2700 times less than the SSH interface, and requires $\delta \approx 55J$.

\section{Robustness to chiral-preserving and chiral-breaking perturbations}
\label{sec:S6}

The OAM interface states require the two zero modes to remain degenerate (main text, Sec.~II~D). Any perturbation that keeps the block-off-diagonal form of Sec.~\ref{sec:S22} guarantees at least $|n_A - n_B| = 2$ zero modes (Sec.~\ref{sec:S23}). For the connectivity used here, nonzero $t_2$ bonds keep the off-diagonal block at full column rank (Sec.~\ref{sec:S22}), so there are exactly two zero modes, and they remain exactly degenerate. A zero mode with a single occupied central charge also persists. At zero energy, the B-site equations of each arm, $\kappa_{+,s}c_+ + \kappa_{-,s}c_- + t_{2,s,1}a_{s,1} = 0$ and $t_{1,s,j-1}a_{s,j-1} + t_{2,s,j}a_{s,j} = 0$, determine all A amplitudes uniquely from $(c_+, c_-)$ whenever every $t_2$ is nonzero, and setting all B amplitudes to zero satisfies the remaining equations. Any central amplitude pair, including $(C, 0)$, therefore extends to an exact zero mode. Its localization requires the arm amplitudes to decay, which for random couplings is controlled by the typical ratio $t_1/t_2$.

\begin{figure}[t]
\centering
\includegraphics[width=0.9\textwidth]{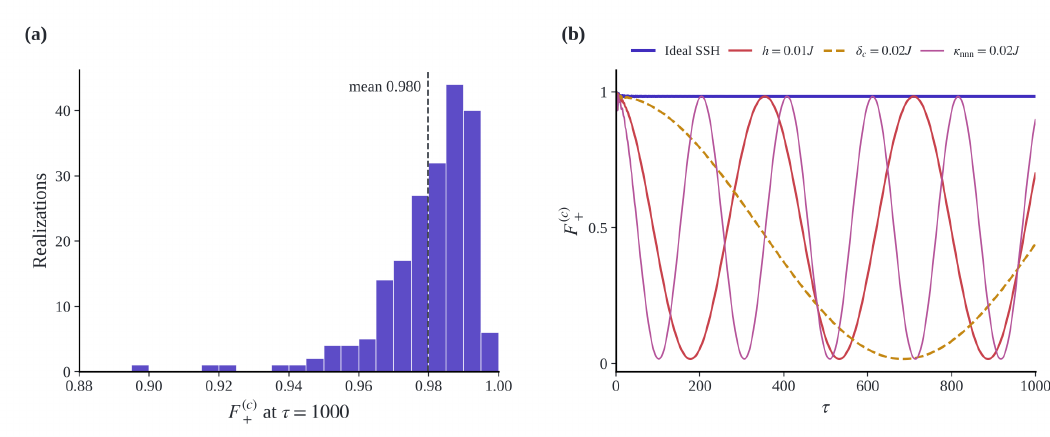}
\caption{Robustness of the OAM interface states. (a) Distribution of $\Fp$ at $\tau = 1000$ for the local $+l$ input over 200 realizations with chiral-preserving disorder: $\pm20\%$ in all arm couplings and $\pm30\%$ in each contact coupling ($M = 500$). The dashed line marks the mean. (b) $\Fp$ versus $\tau$ for the ideal SSH interface and for three chiral-breaking perturbations: direct central mixing $h = 0.01J$, common central detuning $\delta_c = 0.02J$ and central-to-second-neighbor coupling $\kappa_{\mathrm{nnn}} = 0.02J$ ($M = 300$). Other parameters as in Sec.~\ref{sec:S51}.}
\label{fig:S5}
\end{figure}

We test this with 200 realizations of the SSH model of Sec.~\ref{sec:S51} in which every arm coupling is varied independently and uniformly by $\pm20\%$ around its nominal value ($t_2 \in [0.8, 1.2]J$ and $t_1 \in [0.32, 0.48]J$), and each of the four contact couplings $\kappa_{\pm,s}$ is varied independently by $\pm30\%$ around $0.25J$. Each arm contains $M = 500$ dimers, so that dispersive waves do not return to the interface before $\tau = 1000$; for one realization, extending both arms to 600 dimers changes $\Fp(1000)$ by less than $10^{-14}$. In all realizations, the two eigenvalues closest to zero have magnitudes below $2\times10^{-15}J$. For the local $+l$ input, $\Fp$ at $\tau = 1000$ has a mean of 0.980, a median of 0.984 and a minimum of 0.899, and exceeds 0.95 in 96.5\% of the realizations [Fig.~\ref{fig:S5}(a)]; the mean central power is 0.790. With independent contact disorder, a compact central dark state exists only if $\kappa_{+,\mathrm{L}}/\kappa_{-,\mathrm{L}} = \kappa_{+,\mathrm{R}}/\kappa_{-,\mathrm{R}}$, so the bright--dark decomposition of Sec.~\ref{sec:S22} generally no longer applies; the pure-central-charge zero modes follow instead from the recurrence above. The spread of $\Fp$ reflects the changed projection of the local input onto the zero-mode subspace, not a splitting of the doublet.

Perturbations that break chiral symmetry split the doublet. Figure~\ref{fig:S5}(b) shows $\Fp$ for the ideal SSH interface of Sec.~\ref{sec:S51} ($M = 300$, converged against 350 to better than $10^{-13}$) with three separate perturbations: a direct central mixing $h = 0.01J$, a common central detuning $\delta_c = 0.02J$, and a coupling $\kappa_{\mathrm{nnn}} = 0.02J$ between each central orbital and the first A site of each arm. The exact splittings of the two levels nearest the dark-mode propagation constant are $0.017706J$, $0.004588J$ and $0.030813J$, respectively. The first two agree with the perturbative result of Eq.~\eqref{eq:S38}, $(1 + W)h = 0.017706J$ and $(1 - W)\delta_c = 0.004587J$ with $W = 0.770642$. The charge correspondingly beats with $\LB = \pi/\Delta \approx 177/J$, $685/J$ and $102/J$. Retention of a definite charge over a length $L$ therefore requires $L \ll \LB$, which in the physical array is set by the doublet splitting of Sec.~\ref{sec:S32}.

\section{Independent checks of the physical-array calculations}
\label{sec:S7}

The doublet splitting of Sec.~\ref{sec:S32} is the difference of two effective indices that agree to the fifth decimal place, so it is sensitive to the computational domain and mesh. The interface modes decay into the background over about 12\um, which is comparable to the distance between the waveguides and a boundary placed 25\um\ from the array axis. We therefore recomputed the interface modes with an independent full-vectorial finite-difference solver (the Tidy3D mode solver, with subpixel averaging), using the geometry of Sec.~\ref{sec:S32}, perfectly matched layers, and the mirror symmetries of the array. Increasing the half-height of the transverse domain from 25\um\ to 50\um\ changes the splitting by up to $1\times10^{-5}$, whereas a further increase to 75\um\ changes it by less than $2\times10^{-7}$. With a half-height of 50\um\ and uniform mesh steps of 0.5, 0.33, 0.25 and 0.2\um, the splitting is 0.81, 1.4, 0.83 and $1.1\times10^{-5}$ for the $x$-polarized doublet of main-text Fig.~3 and 1.3, 0.93, 1.4 and $1.2\times10^{-5}$ for the $y$-polarized doublet excited in main-text Fig.~4. The values do not approach a limit monotonically; across the tested meshes they lie between $0.8\times10^{-5}$ and $1.4\times10^{-5}$, so we quote the finest-mesh values and regard the magnitude of about $1\times10^{-5}$, rather than the second digit, as established. The MODE values of Sec.~\ref{sec:S32}, 6--$7\times10^{-6}$, lie below this range, consistent with the sensitivity to the domain described above. In the same calculation, the four first-order vector modes of the isolated hollow waveguide are degenerate to within $4\times10^{-6}$, and the effective index of an isolated Gaussian waveguide agrees with the exact step-index value, 1.4401376, to within $2\times10^{-6}$. The splitting of the array doublet therefore arises mainly from the array, as described in Sec.~\ref{sec:S4}, rather than from the vector structure of the hollow-waveguide modes.

We also computed the propagation directly with a scalar split-step Fourier beam-propagation method, which is accurate here because the index contrast is only 0.002. Being scalar, it does not resolve polarization; it checks the propagation in the weak-guidance limit rather than a specific polarized doublet. The transverse window is 480\um\ $\times$ 192\um\ with absorbing edges, the mesh step is 0.5\um\ with subpixel averaging of $n^2$, and the propagation step is 2\um; halving either step changes $\eta_+$ and $P_c$ by less than 0.006. The input is the $l = +1$ superposition of the two first-order modes of the isolated hollow waveguide, obtained by imaginary-distance propagation, so that it occupies only the hollow waveguide, as in main-text Fig.~4. Within the disk $r < 16$\um, which contains the hollow waveguide and its evanescent field, we decompose the field into angular orders $\exp(il\varphi)$ on a polar grid and define $\eta_l$ as the fraction of the power in this disk carried by order $l$, so that the $\eta_l$ sum to one; $P_c$ is the power in the same disk relative to the input. Orders other than $l = \pm1$ carry less than $10^{-4}$ of this power at all distances, so $\eta_+$ equals the conditional fraction $\eta_+/(\eta_+ + \eta_-)$ to within $10^{-4}$. In the isolated hollow waveguide, $\eta_-$ stays below $10^{-4}$ over 60~mm, which confirms that the discretization does not mix the two charges.

\begin{figure}[t]
\centering
\includegraphics[width=0.9\textwidth]{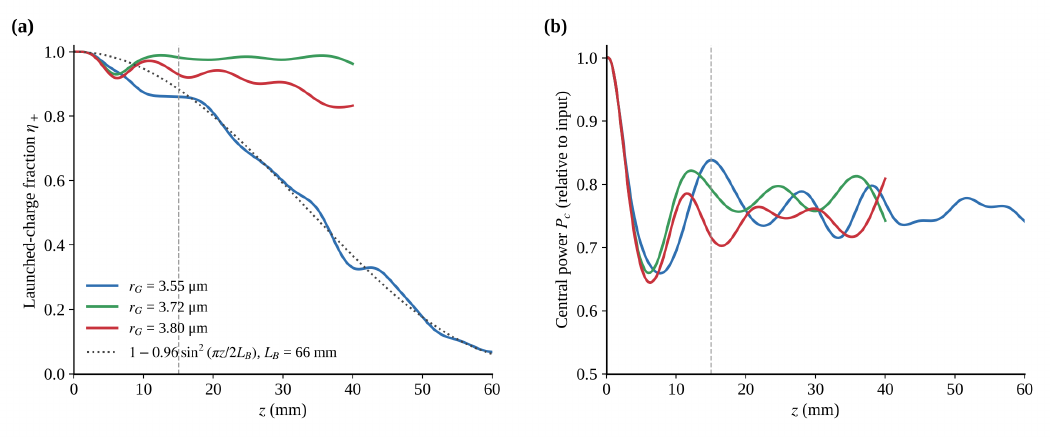}
\caption{Beam-propagation check of OAM retention in the physical array. (a) Fraction $\eta_+$ of the power within $r < 16$\um\ carried by the launched charge, for a local $l = +1$ input and Gaussian-waveguide radii $r_{\mathrm{G}} = 3.55$\um\ (the geometry of main-text Fig.~4), 3.72\um\ and 3.80\um. The dotted line is a fit of the 3.55-$\mu$m curve to $1 - A\sin^2(\pi z/2\LB)$, with $\LB = 66$~mm and $A = 0.96$. (b) Power within $r < 16$\um\ relative to the input. The dashed lines mark $z = 15$~mm.}
\label{fig:S6}
\end{figure}

For the array of main-text Fig.~4 (Gaussian radius $r_{\mathrm{G}} = 3.55$\um), $\eta_+ = 0.97$ at 4~mm and 0.86 at 15~mm, and $P_c$ lies between 0.66 and 0.84 after the initial redistribution (Fig.~\ref{fig:S6}). Over 60~mm, the conversion follows $\eta_- \approx A\sin^2(\pi z/2\LB)$ with a conversion length $\LB = 66$~mm and $A = 0.96$, which corresponds to $\Delta\neff = \lambda/(2\LB) = 1.2\times10^{-5}$ and agrees with the finest-grid mode calculation above. Following Sec.~\ref{sec:S4}, a deliberate mismatch can cancel this splitting. Increasing the Gaussian radius to $r_{\mathrm{G}} = 3.72$\um\ keeps $\eta_+ \geq 0.93$ over 40~mm, with $\eta_+ = 0.98$ at 15~mm; the remaining dips near 6~mm come from the modes other than the interface doublet that the local input excites (Sec.~\ref{sec:S36}). At $r_{\mathrm{G}} = 3.80$\um\ the mismatch overcompensates the splitting, and $\eta_+$ decreases to 0.83 at 40~mm. The full-vectorial mode solver places the zero of the splitting at $r_{\mathrm{G}} \approx 3.70$\um\ for the $x$-polarized doublet and $r_{\mathrm{G}} \approx 3.77$\um\ for the $y$-polarized doublet, consistent with the beam-propagation result.

\renewcommand{\refname}{Supplemental references}